\documentclass{article}
\RequirePackage[OT1]{fontenc}
\usepackage{amsthm,amsmath,amsthm}
\usepackage{graphicx,amsfonts,subcaption,bbm}
\usepackage{rotating,tikz,mathtools,url,lscape,longtable,verbatim}
\usepackage[colorlinks,citecolor=blue,urlcolor=blue]{hyperref}
\usepackage[affil-it]{authblk}

\makeatletter
\renewcommand\AB@affilsepx{\quad \protect\Affilfont}
\makeatother
 
\usepackage[authoryear]{natbib}

\theoremstyle{plain}

\usepackage[margin=1.0in]{geometry}

\begin{document}
\title{\textbf{Bayesian log-Gaussian Cox process regression: with applications to meta-analysis of neuroimaging working memory studies}}
\date{\today}
\author[1]{Pantelis Samartsidis}
\author[2]{Claudia R. Eickhoff}
\author[2]{Simon B. Eickhoff}
\author[3]{Tor D. Wager}
\author[4]{Lisa Feldman Barrett}
\author[5]{Shir Atzil}
\author[6]{Timothy D.\ Johnson}
\author[7]{Thomas E.\ Nichols}

\affil[1]{MRC Biostatistics Unit, University of Cambridge}
\affil[2]{Heinrich-Heine University D\"{u}sseldorf and Forschungszentrum J̈\"{u}lich}
\affil[3]{University of Colorado at Boulder}
\affil[4]{Northeastern University}
\affil[5]{Hebrew University of Jerusalem}
\affil[6]{University of Michigan}
\affil[7]{University of Oxford}

\renewcommand\Authands{ and }
\maketitle

\begin{abstract}
Working memory (WM) was one of the first cognitive processes studied with functional magnetic resonance imaging (fMRI). 
With now over 20 years of studies on WM, each study with tiny sample sizes, the re is a need for meta-analysis to identify the brain regions consistently activated by WM tasks, and to understand the inter-study variation in those activations. 
However, current methods in the field cannot fully account for the spatial nature of neuroimaging meta-analysis data or the heterogeneity observed among WM studies. In this work, we propose a fully Bayesian random-effects meta-regression model based on log-Gaussian Cox processes, which can be used for meta-analysis of neuroimaging studies. An efficient MCMC scheme for posterior simulations is presented which makes use of some recent advances in parallel computing using graphics processing units (GPUs). Application of the proposed model to a real dataset provides valuable insights regarding the function of the WM.      
\end{abstract}

\section{Introduction}

\subsection{The working memory}
Humans depend on working memory (WM) for many behaviours and cognitive tasks. 
WM includes both the retention of information (aka short term memory), as well as the manipulation of information over a short duration. 
An example of the former is remembering a phone number until you dial it, while an example of the latter is building a `mental map' while receiving directions. 
WM is impaired in a number of neurological and psychiatric diseases, most notably in all forms of dementia.  

With its central role in everyday behaviour and implication in disease, WM has been frequently studied with functional brain imaging techniques like functional magnetic resonance imaging (fMRI). 
fMRI is sensitive to changes in blood flow, volume and oxygenation level in the brain, and provides a noninvasive way to identify regions of the brain associated with a given task or behaviour. 
However, each fMRI study has traditionally had very small samples, rarely exceeding 20. 
Thus, there is a need for meta-analysis methods to pool information over studies, separating consistent findings from those occurring by chance, as well as meta-regression methods \citep{Greenland1994} to understand heterogeneity in terms of study-specific characteristics.

\subsection{Neuroimaging meta-analyses}
In fMRI there are two broad approaches for meta-analysis. 
When the full statistical images from each study are available, that is effect sizes and associated standard errors for all voxels in the brain, an intensity-based meta-analysis (IBMA) can proceed by means of standard meta-analytic methods (see \citet{Hartung2008} for an overview). 
However, these statistic images (200,000+ voxels) traditionally have not been shared by authors. 
Instead, researchers only publish the $x,y,z$ brain atlas coordinates of the local maxima
in significant regions of the statistic image. 
We call these coordinates the foci (singular focus). 
When only foci are available then a coordinate-based meta-analysis (CBMA) is conducted. 
As can be expected, the transition from full images to the lists of foci involves a heavy loss of information \citep{Salimi2009}. 
However, since the vast majority of researchers rarely provide the full images, CBMA constitutes the main approach for fMRI meta-analysis.

Most work in the field is focused on the so-called \textit{kernel-based} methods such as activation likelihood estimation \citep[ALE]{Turkeltaub2001,Eickhoff2012}, multilevel kernel density analysis \citep[MKDA]{Wager2004,Wager2007} and signed differential mapping \citep[SDM]{Radua2009,Radua2012}. 
Roughly, these methods construct a statistic map as the convolution of the foci\footnote{Precisely, this is a convolution of a Dirac delta function located at each focus with a given kernel.} with 3D spatial kernels, but not exactly correspond to traditional kernel density estimation. 
In particular, these methods give special treatment to foci that appear close in one study, decreasing their influence relative to dispersed points. 
Areas of the map with large values suggest brain regions of consistent activation across studies. 
For statistical inference, the map is thresholded by reference to a Monte Carlo distribution under the null hypothesis of no consistent activation across studies. 
Kernel-based methods are not based on an explicit probabilistic model and hence often lack interpretability. 
Moreover, for some methods it is difficult to obtain standard errors and hence only p-values are reported for each voxel. 
Some of these approaches cannot accept study-level covariates, and thus can't conduct meta-regression, and all are massively univariate in that they have no model of spatial dependence and can make only limited probabilistic statements about sets of voxels.

Recently, some model-based methods
were proposed to address the limitations of kernel-based methods, such as the
Bayesian hierarchical independent Cox cluster process model of \citet{Kang2011}, the Bayesian nonparametric binary regression model of \citet{Yue2012}, the hierarchical Poisson/Gamma random field model of \citet{Kang2014} and the spatial Bayesian latent factor model of \citet{Montagna2017}. 
However, most of these methods do not allow for meta-regression. 
Further, current model-based approaches do not account for dependence induced when a single publication reports the results of multiple studies using the same cohort of participants. (In this work, we refer to `study' as the result of one statistical map; typically a publication will report results from several maps).

\subsection{Contribution and outline} 
The contributions of this work are twofold. 
The first contribution is methodological. 
In particular, we propose a Bayesian spatial point process model, extension of the log-Gaussian Cox process model \citep{Moller1998} that can account for study specific characteristics as explanatory variables thus allowing for meta-regression. 
Compared to the model of \citet{Montagna2017}, which is the only existing coordinate-based meta-regression method, our model has two advantages. 
Firstly, it is less mathematically complex and therefore easier to communicate to practitioners and elicit prior distributions for its parameters. 
Secondly, by introducing random-effect terms, our model can capture heterogeneities that cannot be captured by the covariates and also reduce biases caused by the assumption that studies in the meta-analysis are independent one of another. 

The second contribution of this paper is to conduct a meta-analysis of working memory fMRI studies using the proposed model. 
Even though previous meta-analyses of working memory studies exist \citep{Wager2003,Owen2005,Rottschy2012}, none of these studies uses some of the available model-based methods and hence the inferences they provide are limited. 
Further, our analyses quantifies the effect of some important covariates and thus provides new insights regarding the function of working memory.

The remainder of this manuscript is structured as follows. 
In Section \ref{sec:dataintro} we present the data under investigation and state the questions that our meta-analysis wishes to answer. 
Motivated by the data in \ref{sec:dataintro}, we introduce our LGCP model in Section \ref{sec:model}. 
The algorithm used for posterior inference is presented in Section \ref{sec:algorithm}. 
The results of the real-data analysis can be found in Section \ref{sec:data}. 
Finally, Section \ref{sec:discussion} summarises our findings and sets some possible directions for future research.

\section{Motivating dataset}\label{sec:dataintro}
\begin{figure}[htp]
		\centering
        \includegraphics[scale=1.2]{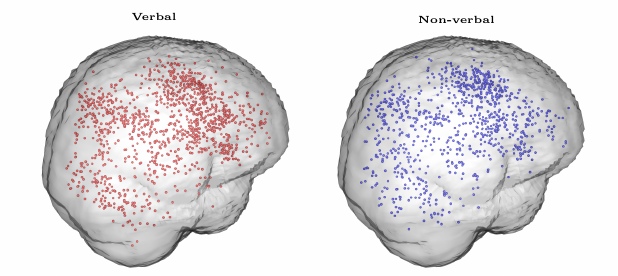}
       \caption{Graphical representation of the meta-analysis dataset. Data consist of 2,107 foci from 157 studies on working memory. Of these, 1,207 are obtained from studies using verbal stimuli (shown in red) whereas the remaining 900 are obtained from studies using non-verbal stimuli (shown in blue). The code used to generate this figure is courtesy of Jian Kang.}
      \label{fig:data}
 \end{figure}

Our investigations are motivated by data from \citet{Rottschy2012}. 
The data have been retrieved from 89 publications on working memory but some of these publications conduct multiple studies (experiments). 
The average number of studies per publication is 1.76 (ranging 1-7). 
Overall, we include 157 studies in the meta-analysis and the total number of foci is 2107. 
As well as the foci, for each study we observe the stimulus type (where 102 studies used verbal stimuli and 55 studies used non-verbal), the sample size (mean 14.94, SD 5.64) and the average age of the participants (mean 32, SD 10.99). 
See Table \ref{tab:data} for more descriptives, whereas a graphical representation of the data can be found in Figure \ref{fig:data}. 
Note that, the dataset that we use is a subset of the dataset of \citet{Rottschy2012}; this is due to missing values for the covariate age. 
 
\begin{longtable}{lcccc}
\caption{Data summaries}
\label{tab:data}\\ 
\multicolumn{5}{c}{\textbf{Dataset composition}} \\ \hline \hline
&Min.\ &Median &Mean& Max.\ \\
Studies per publication & 1 & 1& 1.76 & 7\\
Foci per study & 1 & 11 & 13.42 & 55\\ 
Participants per study & 6 & 14 & 14.94 & 41\\ 
Mean participant age & 21.25 & 29.20 & 32.00 & 75.11\\ \hline
&&&&\\
\multicolumn{5}{c}{\textbf{Verbal}} \\ \hline \hline
&Min.\ &Median &Mean& Max.\ \\
Foci per study& 1 & 10 &11.83 & 39\\
Participants per study& 7 & 14 & 14.91 & 41\\
Mean participant age& 21.80 & 30.12 & 33.80 & 75.11\\ \hline
&&&&\\
\multicolumn{5}{c}{\textbf{Non-verbal}} \\ \hline \hline
&Min.\ &Median &Mean& Max.\ \\
Foci per study& 2 & 15 & 16.36 & 55\\
Participants per study& 6 & 13 & 14.98 & 33\\
Mean participant age& 21.25 & 28.00 & 28.64 & 61\\ \hline
\end{longtable}

Our meta-analysis aims to address the following questions related to the function of working memory: I) what the regions of the brain that are consistently engaged by working memory across studies? II) do these regions differ depending on the type of stimulus presented to the participants? III) is the organisation  of working memory affected by age? IV) does sample size affect the total number of activations reported? 
In order to ensure that the answers to these questions are not driven by influential publications conducting multiple studies, our investigations should account for such dependencies.

\section{A model for CBMA meta-regression}
\label{sec:model}
To address the questions raised in Section \ref{sec:dataintro}, we propose a model for CBMA meta-regression. 
First, we set notation.
Suppose that there are a total $I$ studies in the meta analysis and that each study $i$ comes with a point pattern $\mathbf{x}_i$, a set of foci $x_{ij}\in\mathcal{B}\subset\mathbb{R}^3$, where $\mathcal{B}$ is the support of the analysis, usually set from a standard atlas of the brain, and $j=1,\dots,n_i$, where $n_i$ is the number of foci in a study. 
Additionally, suppose that for each point pattern there is a set of $K$ study specific characteristics, $\left\{z_{ik}\right\}_{k=1}^{K}$. 
Henceforth, we will occasionally refer to these characteristics as covariates.  

We assume that each point pattern $\mathbf{x}_i$ is the realisation of a Cox point process $\mathbf{X}_i$ defined on $\mathcal{B}$, driven by a random intensity $\lambda_i(\cdot)$. 
We can then model the intensity function at each point $\xi\in\mathcal{B}$ as
\begin{eqnarray}\label{intensity} 
\lambda_{i}(\xi) &= &\alpha_i\exp{\left\{ \sum_{k=0}^{K^\ast}{\beta_k(\xi)z_{ik}}+\sum_{k=K^\ast+1}^{K}{\beta_kz_{ik}}  \right\}},
\end{eqnarray}
where $\alpha_i$ is the random effect of study $i$, $\beta_k(\cdot)$ are the regression coefficients for the covariates that have a local effect ($k=0,\dots,K^\ast$), $z_{ik}$ are covariate values where $k=0$ is for the intercept ($z_{i0}=1$), and are $\beta_k$ the regression coefficients for the covariates that have a global (homogenous) effect ($k=K^\ast+1,\dots,K$).

Equation \eqref{intensity} defines a spatial log-linear model over the brain. 
Foci are more likely to occur in regions of the brain with high intensity values whereas we expect almost no foci in regions as the intensity approaches zero. 
The exact rates are given by the properties of a Cox process. 
In particular, given $\lambda_i(\cdot)$, the expected number of foci in any bounded $B\subseteq\mathcal{B}$ is a Poisson random variable with mean $\int_{B}\lambda_i(\xi)d\xi$ \citep{Moller2004}. 

The inclusion of the random effect terms is an important feature of our model. 
Firstly, by assuming that $\alpha_i=\alpha_j$ for studies $i$ and $j$ retrieved from the same publication, we relax the assumption of independence between their reported activations. 
This assumption is taken by all existing CBMA approaches but is unlikely to hold for studies from the same publication. 
For example, a multi-study publication will typically engage the same participants in all of its experiments. 
By using a common random effect for studies from the same publication, our model prevents publications with several studies to drive the estimates of the regression coefficients. 
Secondly, the random effects can allow for additional variability in the total number of foci that cannot be captured by the Poisson log-linear model. 
In a recent study, \citet{PBIAS} found that CBMA data do show such overdispersion and thus inclusion of the random effect terms can potentially improve the fit to the data.  
  
Separation of the covariates into those with a localised and those with a global effect should be done with caution. 
If one is interested in investigating whether the effect of a covariate varies from one region of the brain to another, such as age in our application, a spatially varying regression coefficient is needed. 
However, the total number of parameters associated with a spatially varying effect is large and therefore assigning a spatially varying coefficient to a covariate with a global effect may substantially increase the uncertainty associated to the other model parameters. 
In order to determine if a spatially varying coefficient for a covariate is required, one can fit two models, one that assumes that the covariate has a global effect and one that assumes a local effect. 
If the more complex model improves the fit to the data substantially\footnote{As determined by a goodness-of-fit measure, e.g.\ posterior predictive checks \citep{Gelman1996,Leininger2017}.}, then it should be preferred for inference instead of the simple model. 
Sometimes, it plausible to assume a global effect solely based on prior expectation. For instance, a covariate for multiple testing correction can be assumed to have a global effect; for studies not applying any corrections, we expect false positives to appear uniformly across the brain.

A Bayesian model is defined with prior distributions on model parameters, which here include the functional parameters $\beta_k(\cdot)$ ($k=0,\ldots,K^\ast$), and scalar parameters $\beta_k$, ($k=K^\ast+1,\ldots,K$). 
A natural way to proceed is to assume that $\beta_k(\cdot)$ are realisations of Gaussian processes and that the $\beta_k$ have normal distributions. 
That way, when $\alpha_i=1$, the right hand side of Equation (\ref{intensity}) is also a Gaussian process, and each point process is a log-Gaussian Cox process (LGCP) \citep{Moller1998}. 
The log-Gaussian Cox process is a flexible model for spatial point data that can account for aggregation \citep{Moller1998,Moller2007} or even repulsion between points \citep{Illian2012} and has therefore found applications in several fields such as disease mapping \citep{Benes2002,Liang2009} and ecology \citep{Moller2003,Illian2012b}.

By the definition of a Cox process, $\mathbf{X}_i$ is a Poisson point process on $\mathcal{B}$ conditional on $\lambda_i(\cdot)$ \citep{Moller2004}. The density (Radon-Nikodym derivative) of this point process with respect to the unit rate Poisson process is
\begin{equation}
\pi\left(\mathbf{x}_i|\lambda_i\right)=\exp{\left\{|\mathcal{B}|-\int_{\mathcal{B}}{\lambda_i(\xi)d\xi}\right\}}\prod_{x_{ij}\in\mathbf{x}_i}\lambda_i(x_{ij}),
\end{equation}
for $i=1,\dots,I$, with $|\mathcal{B}|$ denoting the volume of the brain. We can view $\pi\left(\mathbf{x}_i\mid\lambda_i\right)$ as the density of the sampling distribution of the data. 
If we further assume independent studies, then we posterior distribution of the model parameters conditional on the foci is given, up to a normalising constant by 
\begin{multline}\label{lgcppost1}
\pi\left(
\left\{\alpha_i\right\}_{i=1}^{I},
\left\{\beta_k(\cdot)\right\}_{k=0}^{K^\ast},
\left\{\beta_k\right\}_{k=K^\ast+1}^{K}\mid \left\{\mathbf{x}_i\right\}_{i=1}^{I}
\right) \propto
\prod_{i=1}^{I}{\pi(\mathbf{x}_i|\lambda_i)}
\\
\times
\prod_{i=1}^{I}{\pi(\alpha_i)}
\prod_{k=1}^{K^\ast}{\pi(\beta_k(\cdot))}
\prod_{k=K^\ast+1}^{K}{\pi(\beta_k)},
\end{multline}
where $\pi(\alpha_i)$, $\pi(\beta_k(\cdot))$ and $\pi(\beta_k)$ are the priors on the random effects, functional and scalar parameters, respectively, which we discuss the priors below in Section \ref{sec:approximation}.

\subsection{Choice of correlation function}\label{sec:correlation}
We will assume an isotropic, Gaussian correlation structure, that is for points $\xi,\xi'\in\mathcal{B}$ we have
\begin{equation}
\mathbb{C}\mathrm{or}\left(\beta_k(\xi),\beta_k(\xi')\right) = \exp{\left\{-\rho_k||\xi-\xi'||^{\delta_k}\right\}},
\end{equation}
where $\rho_k>0$ are the correlation decay parameters and $\delta_k=2$ for all $k=1,\dots,K^\ast$. 
Note that for numerical stability with the discrete Fourier transform (see Section \ref{sec:algorithm}) we set $\delta=1.9$ in our implementations. 
The same correlation structure was used by \citet{Moller1998} and \citet{Moller2003} in the context of LGCPs.

A Gaussian correlation function is used instead of alternative  correlation structures (see e.g.\ \citet{Rasmussen2005}) because it allows us to calculate the gradient of the posterior with respect to the correlation parameters $\rho_k$, which we use to design an efficient algorithm for posterior simulations (see Section \ref{sec:algorithm} for details). 
Further, in exploratory work using other correlation structures, our neuroscientist colleagues preferred the appearance of results from Gaussian correlation, perhaps because of the pervasive use of Gaussian kernel smoothing in fMRI. 
Finally, it is well known that estimating the correlation parameters for more flexible correlation structures can be extremely challenging in practice, see e.g.\ discussions by \citet{Zhang2004} and \citet{Diggle2013lgcp} for the Mat{\'e}rn correlation function.

\subsection{Posterior approximation}\label{sec:approximation}
Calculation of the posterior in Equation \eqref{lgcppost1} requires the evaluation of the infinite dimensional Gaussian processes $\beta_k\left(\cdot\right)$, $k=0,\dots,K^\ast$, which we approximate with a finite dimensional distribution. 
Following \citet{Moller1998} and \citet{Benes2002}, we consider the discretisation of the 3D volume with a regular rectangular grid $W\supset\mathcal{B}$. 
We use $V$ cubic cells (i.e.\ voxels) in $W$ with volume $A=a^3$, where $a$ is the length of the side. 
In neuroimaging, analysis with $2\mathrm{mm}^3$ cubic voxels is typical, leading to a box-shaper grid of about 1 million voxels, of which about 200,000 are in the brain or cerebellum. 
Note that for simplicity, we consider both grey matter and white matter voxels in our implementations. 
Voxels are indexed $v=1,\ldots,V$, and the coordinate of $v$ is the location of the center $\boldsymbol{\nu}_v\in \mathbb{R}^3$.

For any $k=0,\dots,K^\ast$, the Gaussian process $\beta_k(\cdot)$ can be now approximated with a  step function which is constant within each voxel $v$ and equal to the value of $\beta_k(\cdot)$ at the location of the center, i.e.\ $\beta_k(\boldsymbol\nu_v)$. 
\citet{Waagepetersen2004} shows that the accuracy of this approximation improves as $a$ goes to zero. By definition, $\boldsymbol\beta_k=\left[\beta_k(\boldsymbol\nu_1),\dots ,\beta_k(\boldsymbol\nu_V)\right]$ are multivariate Gaussian vectors. 
We parametrise $\boldsymbol\beta_k$ as
\begin{equation}\label{lgcpparam}
\boldsymbol\beta_k=\mu_k\mathbf{1}_V+\sigma_k \mathbf{R}_k^{1/2}\boldsymbol\gamma_k,
\end{equation}
where $\mu_k$ are the overall (scalar) means, $\mathbf{1}_V$ is a $V$-vector of ones, $\sigma_k$ are the marginal standard deviations, $\mathbf{R}_k$ are the $V\times V$ correlation matrices with elements $\left(\mathbf{R}_k\right)_{ij}=\exp{\left\{-\rho_k||\boldsymbol\nu_i,\boldsymbol\nu_j||^{2}\right\}}$, and $\boldsymbol\gamma_k$ are the \textit{a priori} $\mathcal{N}_V(\mathbf{0},\mathbf{I}_{V})$ vectors, $k=0,\dots,K^\ast$. 
The same parametrisation is used by \citet{Moller1998}, \citet{Christensen2002} and is advocated by \citet{Christensen2006} because it allows for computationally efficient posterior simulations.

Priors for the $V$-vectors $\boldsymbol\gamma_k$ are induced by the parametrisation of Equation \eqref{lgcpparam}. 
The priors for the remaining model parameters are set as follows. 
We assign weakly informative $\mathcal{N}\left(0,10^8\right)$ to the scalar parameters $\mu_k$, $\sigma_k$ and $\beta_k$.  
Further, we assume that $\rho_k\sim\mathrm{Uni}\left[3.5\times10^{-3},0.1\right]$, which we found corresponded to smoothness ranges found in single-study fMRI statistic maps. 
Finally, in order to ensure identifiability, we \textit{a priori} let $\alpha_i\sim\mathcal{G}(\kappa,\kappa)$. 
In our analyses, we set $\kappa=10$ since we expect 90\% of the multiplicative random effects to be within the interval $[0.5,1.5]$.

Once the latent Gaussian processes are approximated, one can also approximate  $\lambda_i$ with a step function as before. 
The intensities at the center of each voxel are given by
\begin{equation}\label{eq:lgcpipost}
\boldsymbol\lambda_i=
\alpha_i
\exp{\left\{ \sum_{k=0}^{K^\ast}{\left(\mu_k\mathbf{1}_V+\sigma_k\mathbf{R}^{1/2}_k\boldsymbol\gamma_k\right)z_{ik}} +\sum_{k=K^\ast+1}^{K}{\beta_kz_{ik}}\mathbf{1}_V \right\}},
\end{equation}
where $\boldsymbol{\lambda}_i$ is the $V$-vector, the discretised intensity. 
We will write $\lambda_{iv}=(\boldsymbol{\lambda}_i)_v$ for the $v$-element of study $i$'s intensity. 
The approximated posterior is
\begin{equation}\label{eq:lgcptpost}
\pi\left(\boldsymbol\theta\mid \left\{\mathbf{x}_i\right\}_{i=1}^{I}\right)\propto \prod_{i=1}^{I}{\left[\exp{\left\{-\sum_{v}{A_v\lambda_{iv}}\right\}}\prod_{j=1}^{n_i}{\lambda_{iv(x_{ij})}}\right]}\pi(\boldsymbol\theta),
\end{equation}
where $\boldsymbol\theta=\left\{  \left\{\alpha_i\right\}_{i=1}^I,\left\{\mu_k\right\}_{k=1}^{K^\ast},\left\{\sigma_k\right\}_{k=1}^{K^\ast},\left\{\rho_k\right\}_{k=1}^{K^\ast},\left\{\boldsymbol\gamma_k\right\}_{k=1}^{K^\ast},\left\{\beta_k\right\}_{k=K^\ast+1}^{K}\right\}$, $A_v$ takes on the value $A$ when $\boldsymbol{\nu}_v \in \mathcal{B}$ and 0 otherwise, $v(x_{ij})$ is the index of the voxel containing $x_{ij}$, and $\pi(\boldsymbol\theta)$ is the joint prior distribution of the parameters. 
The posterior distribution in Equation \eqref{eq:lgcptpost} is still analytically intractable due to the presence of an unknown normalising constant and thus we need to resort to Monte Carlo simulation or approximation techniques to obtain samples from it. 
The method that we use is described in Section \ref{sec:algorithm}.

\section{Sampling algorithm details}\label{sec:algorithm}

Bayesian methodology for inference on LGCPs can be broadly divided into two main categories:
simulation based approximations of the posterior such as Markov chain Monte Carlo \citep[MCMC]{Moller1998} and elliptical slice sampling \citep{Murray2010}, and deterministic approximations to the posterior such as integrated nested Laplace approximations \citep[INLA]{Illian2012,Simpson2011} and variational Bayes \citep{Jaakkola2000}. 
In a recent study, \citet{Taylor2014} compare the Metropolis-adjusted Langevin (MALA) algorithm with INLA and find that both methods give similar results. 
In our application, we choose to use simulation based methods because application on our 3D problem is more straightforward.  

We propose a hybrid MCMC algorithm to sample from the posterior \eqref{eq:lgcptpost}, where parameters are updated in two blocks. 
The first block includes the random effect terms $\boldsymbol{\alpha}=\left\{\alpha_i\right\}_{i=1}^{I}$, whereas the second block includes the remaining model parameters $\boldsymbol{\theta}^\ast=\boldsymbol{\theta}\setminus\boldsymbol{\alpha}$. 
The gamma prior is conjugate for the elements of $\boldsymbol{\alpha}$; hence, they are simulated from their full conditional distributions given the remaining model parameters, see Appendix \ref{sec:rfxgibbs} for details. 
Even though it is possible, we choose not to update $\boldsymbol{\alpha}$ jointly with $\boldsymbol{\theta}^\ast$ because that would increase computation time of our algorithm.

Sampling from the full conditional of $\boldsymbol{\theta}^\ast$ given $\boldsymbol{\alpha}$ is challenging due to its dimensionality. 
\citet{Girolami2011} showed that of all possible strategies, their Riemann manifold Hamiltonian Monte Carlo (RMHMC) sampler is the computationally most efficient for LGCPs in a 2D setting. 
Unfortunately, application in this problem (3D setting) is prohibitive as it would require the inversion of a huge $V\times V$ tensor matrix. 
Alternatives to RMHMC include the MALA and the standard Hamiltonian Monte Carlo \citep[HMC]{Duane1987,Neal2011} algorithms. 
We choose to use HMC because \citet{Girolami2011} found that it is more efficient compared MALA in a 2D setting. 
This finding was confirmed in our preliminary 2D simulation studies with synthetic CBMA data, where HMC outperformed MALA in terms of computational efficiency (mixing/running-time tradeoff).

HMC initially appeared in the physics literature by \citet{Duane1987} under the name \textit{hybrid Monte Carlo}, and later emerged into statistics literature by \citet{Neal2011}.  
HMC emulates the evolution of a particle system which is characterised by its position ($\mathbf{q}$) and momentum ($\mathbf{p}$) over time. 
In our case, $\mathbf{q}$ will be the parameter vector of interest $\boldsymbol{\theta}^\ast$, and $\mathbf{p}$ will be introduced artificially from a $\mathcal{N}_d(0,\mathbf{M})$ distribution, with $d$ being the dimensionality of the problem and $\mathbf{M}$ the mass matrix. 
The dynamics of the system are described by a set of differential equations, known as Hamilton's equations.
 
HMC alternates between moves for the position vector $\boldsymbol\theta^\ast$ and the momentum vector $\mathbf{p}$ based on Hamilton's equations. 
If the solutions of the equations can be found analytically then moves will be deterministic; if not, numerical integration is required and an acceptance/rejection step must be performed to account for integration error. 
Integration is done in fictitious time $\epsilon L$, where $\epsilon$ is the \textit{stepsize} and $L$ is the \textit{number of steps}. 
Typically the \textit{leapfrog integrator} is employed, which for $L=1$ and starting at time $t$ is performed as \citep{Neal2011}
\begin{eqnarray}\label{leapfrog}\nonumber
\mathbf{p}\left(t+\frac{\epsilon}{2}\right)&=&\mathbf{p}\left(t\right)+\frac{\epsilon}{2}\nabla_{\boldsymbol\theta^\ast}\log{\pi\left(\boldsymbol\theta^\ast(t)\mid \left\{\mathbf{x}_i\right\}_{i=1}^{I},\boldsymbol{\alpha}\right)}\\
\boldsymbol\theta^\ast\left(t+\epsilon\right)&=&\boldsymbol\theta^\ast\left(t\right)+\epsilon \mathbf{M}^{-1}\mathbf{p}\left(t+\frac{\epsilon}{2}\right)\\ \nonumber
\mathbf{p}\left(t+\epsilon\right)&=&\mathbf{p}\left(t+\frac{\epsilon}{2}\right)+\frac{\epsilon}{2}\nabla_{\boldsymbol\theta^\ast}\log{\pi\left(\boldsymbol\theta^\ast(t+\epsilon)\mid \left\{\mathbf{x}_i\right\}_{i=1}^{I},\boldsymbol{\alpha}\right)}.
\end{eqnarray}

Overall, if the method is applied correctly, it will produce samples from the desired full conditional distribution $\pi\left(\boldsymbol\theta^\ast\mid \left\{\mathbf{x}_i\right\}_{i=1}^{I},\boldsymbol{\alpha}\right)$. 
Gradient expressions for the elements of $\boldsymbol{\theta}^\ast$, including correlation parameters $\rho_k$, can be found in Appendix \ref{sec:lgcpgrad}. 
Since it is well known that grouping of variables can lead to samplers with faster convergence properties \citep{Park2009}, we choose to update all elements of $\boldsymbol{\theta}^\ast$ jointly using the HMC. 
The solutions to Hamilton's equations are not available analytically so we need to use the Leapfrog integrator and include an accept/reject step at the end of it.  

Our sampler requires the specification of a stepsize $\epsilon$ and a total number of leapfrog steps $L$ for the HMC step.
\citet{Hoffman2014} show how tuning can be achieved automatically but when we applied this method to our problem running time was increased substantially. 
Therefore we use an alternative approach to tune these parameters. 
The stepsize is automatically adjusted during the burn-in phase of the HMC to give an overall acceptance rate close to the $65\%$ suggested by \citet{Neal2011}. 
In particular, if $\epsilon_t$ is the stepsize at iteration $t$ and $q_{t_1}$ is the acceptance rate over the past $t_1$ iterations, then every $t_2$ iterations we calculate the new stepsize $\epsilon_t'$ as
\begin{equation}
\epsilon_t'=\begin{cases}
0.9\epsilon_t & q_{t_1}<0.60\\
\epsilon_t & 0.60\leq q_{t_1}\leq 0.70\\
1.1\epsilon_t & q_{t_1}>0.70
\end{cases}.
\end{equation}
Specifically we use $t_1=100$ and $t_2=10$.
A similar approach is employed by \citet{Marshall2012} for MALA. 
The latter (number of leapfrog steps), is always fixed to $L=50$. 
We took this approach because we found that, for our LGCP application, the mixing properties of the algorithm scale linearly with $L$ but also with the total number of HMC iterations. 
Hence one can use a relatively large $L$ and few iterations or relatively smaller $L$ and more iterations, the total computation time staying relatively constant.

The last tuning parameter in the HMC algorithm is the variance-covariance matrix of the zero mean normal momentum parameters, $\mathbf{M}$. 
To our knowledge, there is only limited off the shelf methodology on how to adjust $\mathbf{M}$. 
As a starting place we set $\mathbf{M}=\mathbf{I}$. 
\citet{Neal1996} suggests that if an estimate of the posterior variance $\hat{\boldsymbol\Sigma}_{\boldsymbol\theta^\ast}$ is available then a good practice is to set $\mathbf{M}=\hat{\boldsymbol\Sigma}_{\boldsymbol\theta^\ast}^{-1}$. 
In principle, $\hat{\boldsymbol\Sigma}_{\boldsymbol\theta^\ast}$ can be estimated during the burn-in phase of HMC but in practice this is not possible due to the dimensionality of the problem. 
In our simulations, we found that the mean posterior variance of the elements of the $\boldsymbol\gamma_k$ was higher compared to the scalar parameters, followed by $\beta_k$ or $\sigma_k$ and then $\rho_k$. 
Especially for the $\rho_k$ the scale is typically much smaller compared to the other parameters in our applications and so we use $100\times\rho_k$ instead of $\rho_k$. 
After the reparametrisation we found that setting the mass for parameters of $\boldsymbol\gamma_k$, $\beta_k$, $\sigma_k$ and $\rho_k$ equal to 1, 9, 16 and 25 respectively worked well in most of our implementations on simulated and real data. 
However, users might need to adjust these parameters if mixing of the chains is slow. 
For example, estimates of the posterior variance of the scalar parameters can be obtained based on preliminary runs of the algorithm for a few iterations. 
In Appendix \ref{sec:simulations}, we perform a series of simulations studies which demonstrate that the proposed HMC algorithm can efficiently sample from the posterior distribution of the high-dimensional parameter vector $\boldsymbol{\theta}^\ast$.

The most computationally demanding part of the algorithm is the  the calculation of the large matrix-vector products $\mathbf{R}^{1/2}_k\boldsymbol\gamma_k$ appearing in 
the intensity functions of Equation \eqref{eq:lgcpipost}. 
Luckily, an elegant solution to this problem is given by \citet{Moller1998} based on \textit{circulant embedding} that was first proposed by \citet{Dietrich1993} and \citet{Wood1994}. 
The key to the approach is the linear algebra result that a circulant matrix has the discrete Fourier basis as its eigenvectors. 
$\mathbf{R_k}$ is not circulant but is block Toeplitz and can be embedded in a $(2V)\times(2V)$ matrix that is circulant. 
Thus the matrix square root, inversion and multiplication can be accelerated by using (the highly efficient) discrete Fourier transform (DFT) of the embedded matrix and manipulating Fourier coefficients, followed by inverse DFT and extracting the appropriate sub-matrix/sub-vector. 
See \citet[Section 2.6.2]{Rue2005} for more details.

We close this section by stressing that despite the massive dimensionality of the parameter vector, the problem has a very high degree of parallelisation. 
Intensities can be evaluated in blocks of thousands of voxels simultaneously making the algorithm suitable for implementation in a \textit{graphics processing unit} (GPU). 
The most computationally intensive part of our model, namely  operations with DFTs, is also amenable to parallelisation and there exist libraries such as NVIDIA's cuFFT library that are designed for this specific task. 
Overall, we believe that implementation of the log-Gaussian Cox process model described above will soon become a routine task for any moderately powerful GPU device.


\section{Analysis of the WM dataset}\label{sec:data}
\subsection{Model, algorithm details and convergence diagnostics}
For $i=1,\dots,157$ we fit the model
\begin{equation}\label{eq:reallgcp}
\boldsymbol\lambda_{i}=\alpha_i\exp\Big\{
\boldsymbol{\beta}_{0}d_{i0} + \boldsymbol{\beta}_1d_{i1}+\boldsymbol{\beta}_2\mathrm{age_i}+\beta_3\frac{1}{\sqrt{n_i}}\mathbf{1}_V
\Big\}, 
\end{equation}
where $d_{i0}$ and $d_{i1}$ are indicator variables of verbal and non-verbal stimuli, respectively, and $n_i$ is the total number of participants in study $i$. 
Continuous parameters were standardised before implementation. 

We run the MCMC algorithm described in Section \ref{sec:algorithm} for 22,000 iterations, discarding the first 7,000 as a burn-in. 
The algorithm run for approximately 30 hours on an NVIDIA Tesla K20c GPU card. 
We then apply a thinning factor of 15 to the chains and therefore end up with 1,000 draws from the posterior distribution of the model parameters. 
The total number of leapfrog steps is set to $L=50$ and the stepsize is initialised at $\epsilon=0.00001$. 
We use a diagonal mass matrix with units specified in Section \ref{sec:algorithm}. 
A preliminary run of the algorithm revealed that the posterior variance of the scalar parameters $\rho_2$ and $\sigma_2$ of $\boldsymbol{\beta}_2$ was higher compared to the corresponding parameters of $\boldsymbol{\beta}_0$ and $\boldsymbol{\beta}_1$. 
Therefore, in order to improve mixing of the algorithm, we set the mass parameters to 1 and 4 for $\rho_2$ and $\sigma_2$, respectively.  

Convergence of the MCMC chain is assessed visually by inspection of posterior traceplots for the model parameters. 
We run a total of 2 MCMC chains in order to examine if they all converge to the same values. 
Posterior traceplots are shown in Appendix \ref{sec:datasupp}. 
Due to the large number of parameters we mainly focus on the scalar parameters of the model and some summary statistics, see Appendix \ref{sec:datasupp} for more details. 
Results indicate that our chains have converged to their stationary distribution. 
This is verified by the fact that posterior values from the 2 different runs overlap one with another for all the quantities that we examine. 

\subsection{Results}

\begin{landscape}
\begin{figure}[htp]
		\centering
        \vspace{-0.90in}
        \includegraphics[scale=1.05]{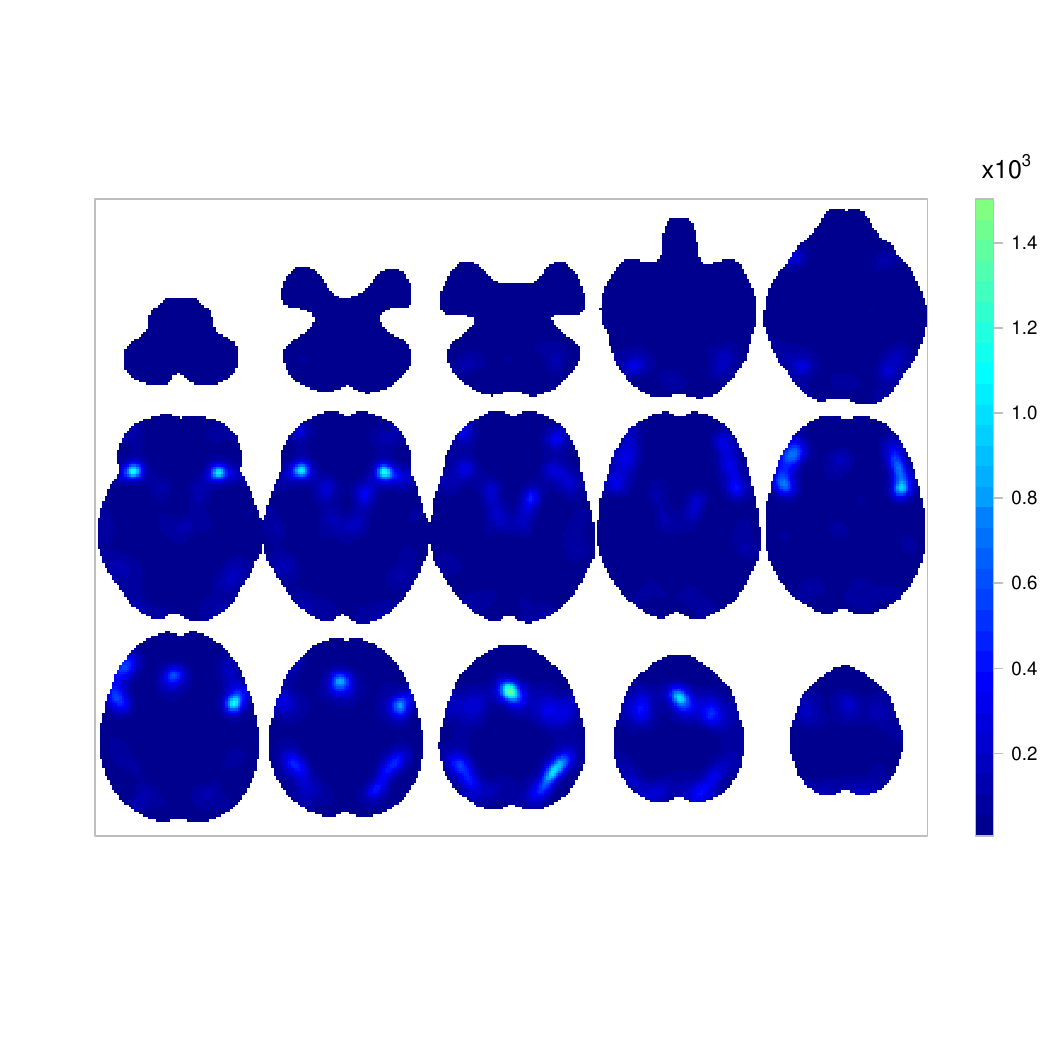}
        \vspace{-1.4in}
       \caption{Voxelwise mean posterior of $\boldsymbol\lambda$, the average intensity of a working memory study. Top row shows (from left to right) axial slices $z=-50,-42,-34,-26$ and $-18$, respectively. Middle row shows axial slices $z=-10,-2,+6,+14$ and $+22$, respectively. Bottom row shows axial slices $z=+30,+38,+46,+54$ and $+62$, respectively.}
      \label{fig:wm}
 \end{figure}
\end{landscape}

Figure \ref{fig:wm} shows the mean posterior of $\boldsymbol\lambda$, the average intensity of a working memory study, where $\boldsymbol\lambda=(\boldsymbol\lambda_{\mathrm{v}}+\boldsymbol\lambda_{\mathrm{nv}})/2$, $\boldsymbol\lambda_{\mathrm{v}}$ is the intensity for verbal WM studies and $\boldsymbol\lambda_\mathrm{nv}$ is for non-verbal WM studies (mean age and  number of participants are set equal equal to the average values in our dataset).  We can see that working memory engages several regions of the brain. 
The regions mostly activated are the frontal orbital cortex (axial slice $z=-10$, left), the insular cortex ($z=-10$, right and $z=-2$, left and right), the precentral gyrus ($z=+30$, left), Broca's areas ($z=+22$ \& $z=+30$, bilateral), the angular gyrus ($z=+46$, left), the superior parietal lobule ($z=+46$, right) and the paracingulate gyrus ($z=+46$, middle). 

Our results are qualitatively similar to results obtained by \citet{Rottschy2012} who used the ALE method. 
However, our model-based approach allows us to derive several quantities of interest along with credible intervals, that cannot be obtained by any of the kernel-based methods. 
For example, one may calculate the probability of observing at least one focus in a set of voxels, e.g.\ an ROI or the entire brain. 
Table \ref{tab:roi} summarises the posterior distribution of $\mathbb{P}(N_\mathbf{X}(B)\geq 1)$, the probability of observing at least one focus in $B$, for several ROIs $B$. 
A full brain analysis can be found in Appendix \ref{app:fullbrain}. 
The division of the brain in ROIs is done according to the Harvard-Oxford atlas \citep{Desikan2006}.

\begin{longtable}{crc|rr}
\caption{Posterior \% probabilities of observing at least once focus for several ROIs. All quantities have been calculated based on 1,000 MCMC samples.}
\label{tab:roi}\\ 
\textbf{ROI} &\textbf{Mean}&\textbf{95\% CI}& \textbf{Verbal}& \textbf{Non-verbal}\\
\hline\hline
Frontal orbital cortex&36.94&[27.27,43.06]&37.26&36.48\\
Insular cortex&33.39&[26.68,39.36]&32.79&33.86\\
Precentral gyrus&68.47&[59.96,73.72]&64.10&72.09\\
Inferior frontal gyrus, PO&39.88&[31.06,45.96]&43.66&35.69\\
Angular gyrus&21.69&[14.39,26.34]&24.30&18.91\\
Superior parietal lobule&36.16&[26.16,42.31]&38.81&33.24\\
Paracingulate gyrus&46.22&[35.94,52.89]&42.91&49.14\\
\hline
\end{longtable} 

\begin{landscape}
\begin{figure}[htp]
		\centering
        \vspace{-0.90in}
        \includegraphics[scale=1.05]{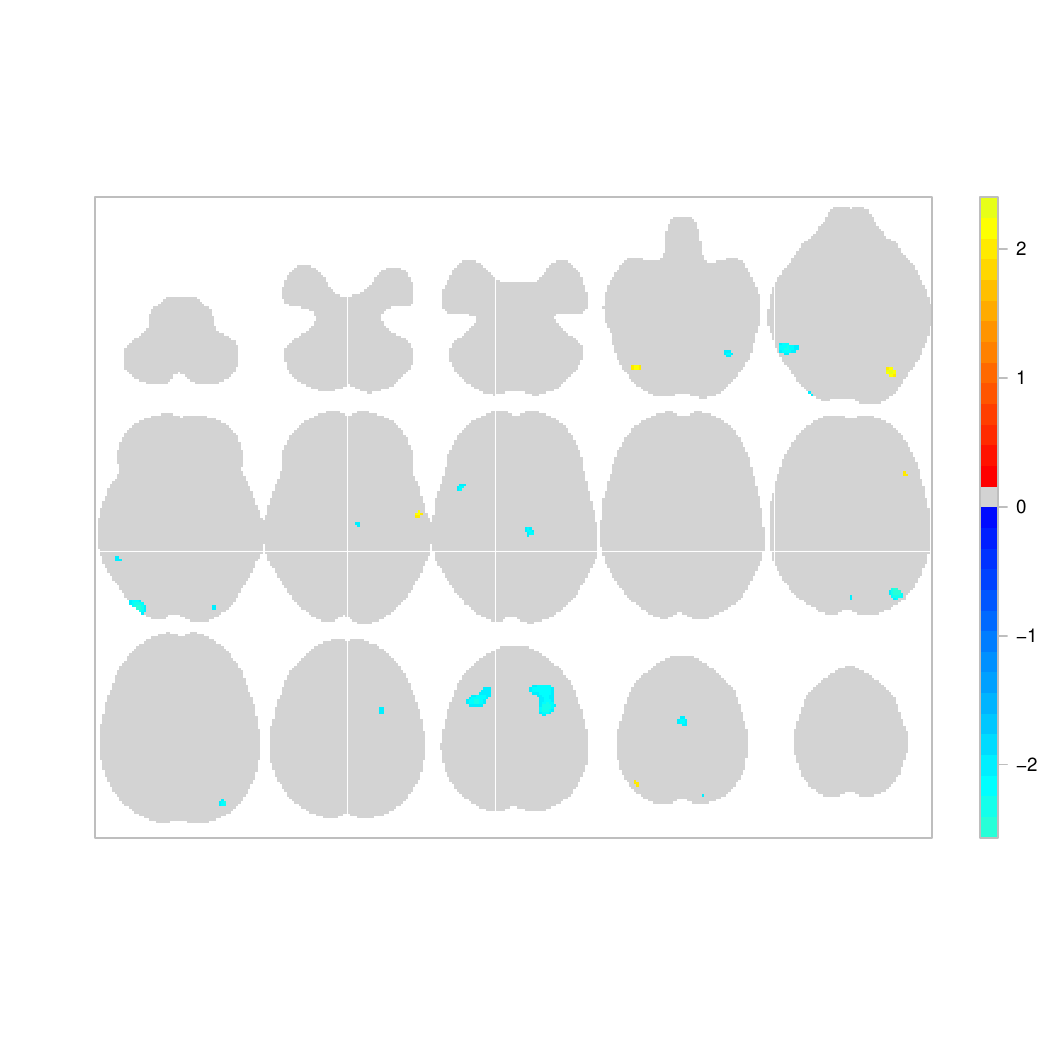}
        \vspace{-1.4in}
       \caption{Voxelwise mean standardised posterior difference between  $\boldsymbol\beta_1$ and $\boldsymbol\beta_2$, the intensities of studies using verbal and non-verbal stimuli, respectively. Top row shows (from left to right) axial slices $z=-50,-42,-34,-26$ and $-18$, respectively. Middle row shows axial slices $z=-10,-2,+6,+14$ and $+22$, respectively. Bottom row shows axial slices $z=+30,+38,+46,+54$ and $+62$, respectively. Voxels for which the mean posterior $\boldsymbol\lambda$ is low (below the 75\% quantile over the brain) or the absolute mean standardised posterior difference is less than two have been set to zero.}
      \label{fig:wmdiff}
 \end{figure}
\end{landscape}

We use posterior intensities $\boldsymbol\lambda_\mathrm{v}$ and $\boldsymbol\lambda_\mathrm{nv}$ to compare activation between the two types of studies in our sample, namely studies using verbal and studies non-verbal stimuli. 
We start with an ROI analysis. 
In particular, for each type and ROI we calculate the probability of at least one focus observed as explained above. 
These are shown in Table \ref{tab:roi} for a few ROIs, whereas a full brain analysis of the two types can be found in Appendix \ref{app:fullbrain}. 
We see that even though the two types show similar patterns of activation, there several ROIs where the probabilities of at least one focus have credible intervals with little overlap. 
The main differences are found in the superior frontal gyrus, the middle frontal gyrus, the lateral occipital cortex, superior division and the inferior frontal gyrus, pars opercularis. 
A voxel-by-voxel comparison is also feasible. 
To answer this, we use the mean standardised posterior difference $\frac{\beta_{0v}-\beta_{1v}}{\mathrm{sd}\left(\beta_{0v}-\beta_{1v}\right)}$. 
This is shown in Figure \ref{fig:wmdiff}. 
Large positive values indicate regions that are activated by verbal stimuli more than non-verbal stimuli. 
Such regions appear the occipital fusiform gyrus ($z=-18$, right).  
Based on the mean standardised posterior difference, regions mostly activated in studies using non-verbal are located in the middle frontal gyrus ($z=+46$).   

\begin{landscape}
\begin{figure}[htp]
		\centering
        \vspace{-0.90in}
        \includegraphics[scale=1.05]{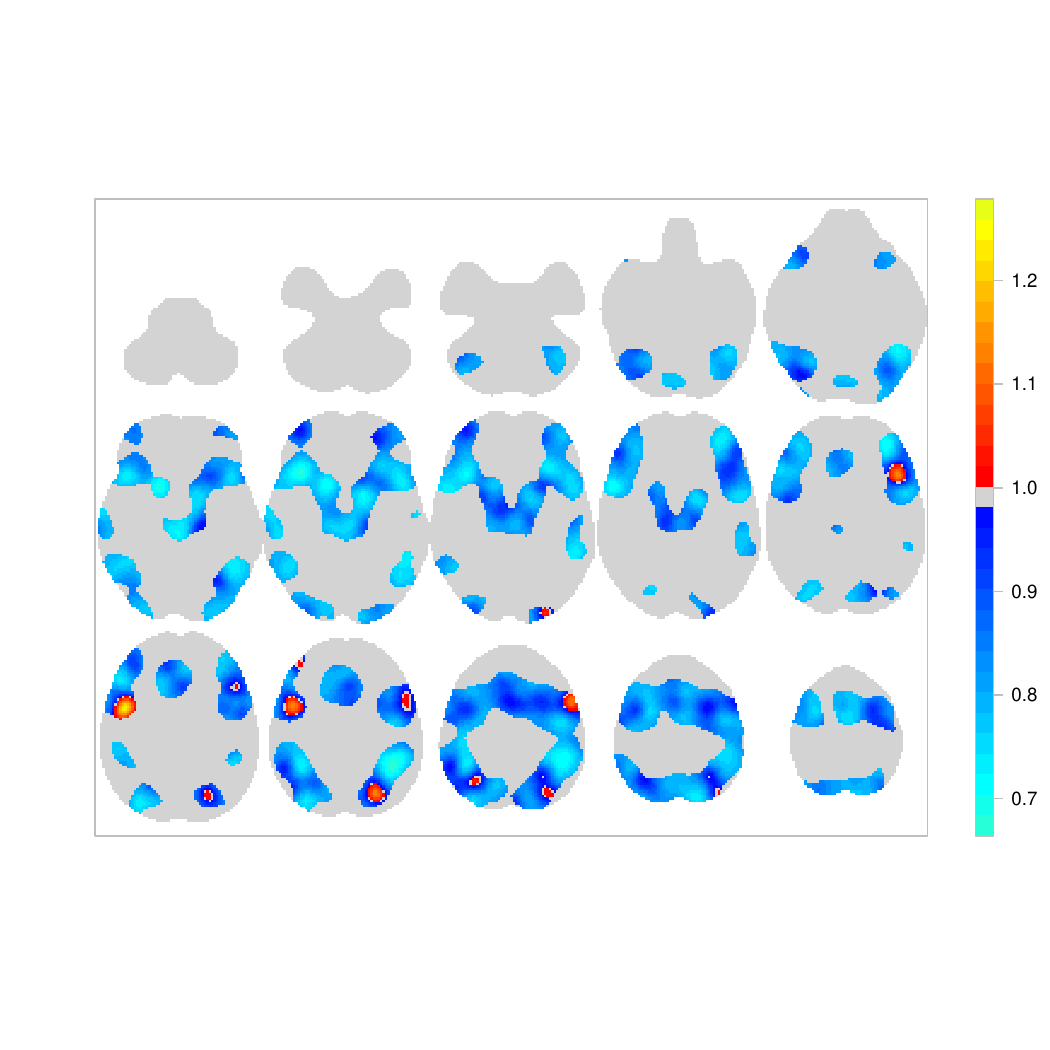}
        \vspace{-1.4in}
       \caption{Mean posterior of $\exp{\left\{\boldsymbol{\beta}_2\right\}}$, the multiplicative age effect on the intensity of both verbal and non-verbal studies. Top row shows (from left to right) axial slices $z=-50,-42,-34,-26$ and $-18$, respectively. Middle row shows axial slices $z=-10,-2,+6,+14$ and $+22$, respectively. Bottom row shows axial slices $z=+30,+38,+46,+54$ and $+62$, respectively. Voxels for which the mean posterior $\boldsymbol\lambda$ is low (below the 75\% quantile over the brain) have been set to one.}
      \label{fig:wmage}
 \end{figure}
\end{landscape}

Our results provide evidence that age has an important effect on the function of working memory. 
The point estimate for the overall age effect $\mu_2$ is -0.22 (95\% CI [-0.337,-0.120]) thus suggesting that we expect a decrease of 20\% in the total number of reported activations per study, each time the average age of the participants increases by 10.99 years. 
Localised age effects can be identified through the posterior distribution of $\exp{\left\{\boldsymbol{\beta}_2\right\}}$, the mean of which is shown in Figure \ref{fig:wmage}. 
The map represents the multiplicative effect that an increase of the average participant age by 10.99 years has on the intensity of both verbal and non-verbal studies. 
Large negative age effects can be found near the left putamen ($z=-2$ and $z=-10$, middle), the insular cortex ($z=-2$, left) and near the superior parietal lobule ($z=+38$ and $z=+46$, right). 
A positive age effect is found near the precentral gyrus ($z=+30$, left). 
However, due to the limited number of studies, the posterior variance of these estimates is large in some regions of the brain, see Figure \ref{fig:agevar} of Appendix \ref{sec:datasupp}. 

The 95\% CI for the sample size covariate is [-0.088,0.064] thus indicating that there is no significant effect on the total number of reported activations. The result is counter-intuitive as one would expect that studies with few participants would be underpowered and thus detect fewer activations. Thus, further investigation is required.

\begin{figure}[h]
		\centering
        \includegraphics[scale=0.5]{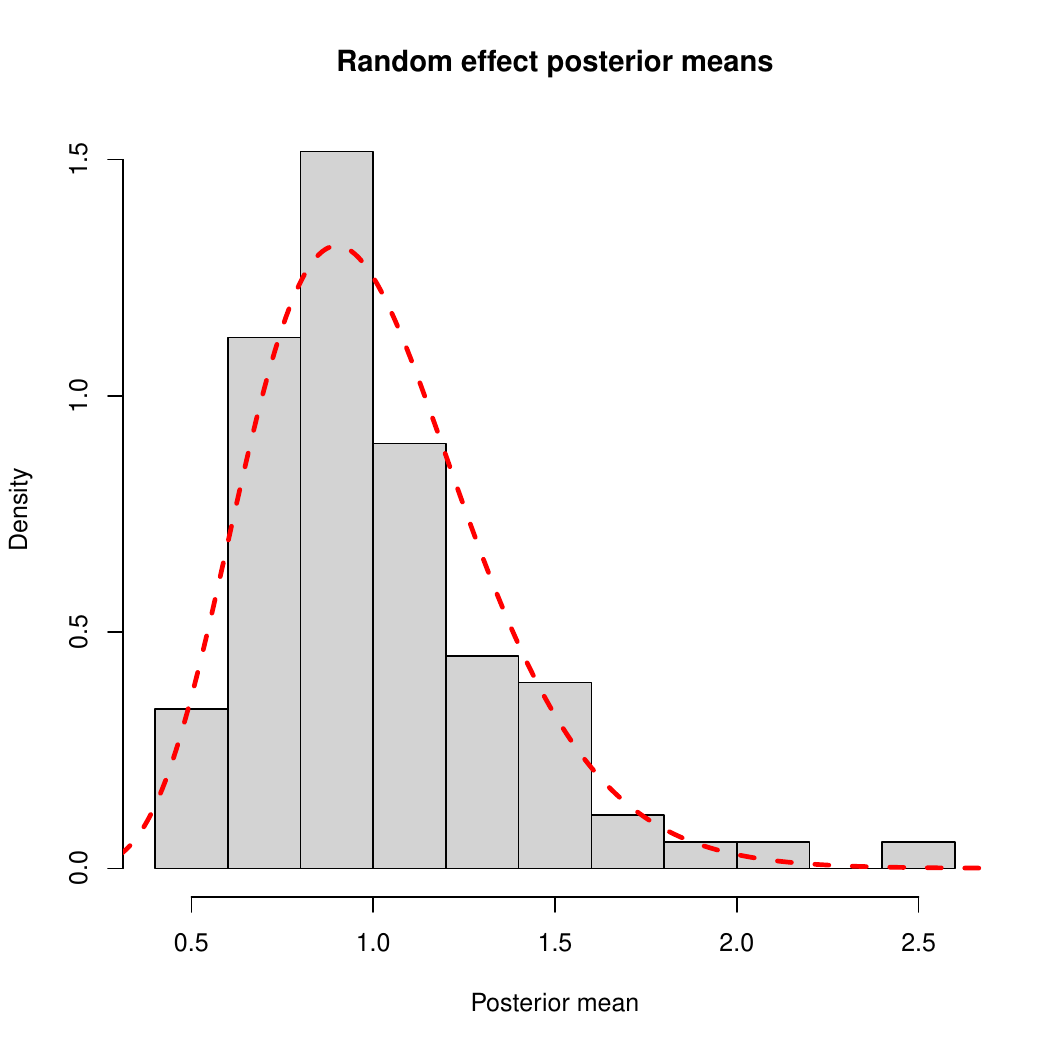}
       \caption{Histogram of the mean posterior random effect terms, $\alpha_i$. We only plot the 89 unique random effects, one for each publication considered in the meta-analysis. Means are based on a sample of 1,000 MCMC draws from the posterior. The dashed red line represents the density of the Gamma prior.}
      \label{fig:rfx}
 \end{figure}

Figure \ref{fig:rfx} shows the mean posterior of the 89 unique random effect terms $\alpha_i$, one for each publication considered. 
We see that despite most of the mass being near 1, there are publications whose mean posterior random effect is different than 1, thus suggesting that observed variability of the foci counts is larger compared to what can be explained by the Poisson log-linear model. 
The importance of allowing for this additional variability can be seen by comparing the proposed random effects model to the standard LGCP model, which we also fit to the data. 
We use posterior predictive checks \citep{Gelman1996} to assess how well the two models fit the data. 
For each study and MCMC draw, we simulate from the posterior predictive distribution of $N_{\mathbf{X}_i}(\mathcal{B})$, the total number of foci, given the covariates. 
Based on these draws, we calculate the 95\% predictive intervals of $N_{\mathbf{X}_i}(\mathcal{B})$ and check if they contain the observed values. 
For our model, the coverage of the intervals is $90\%$ compared to $66\%$ obtained using the standard LGCP model, which implies that our model provides a better fit to the data compared to the standard LGCP.  
A comparison of the predictive intervals that takes into account the length of these intervals can be based on the mean interval score \citep{Gneitning2007}. 
This is 22.45 and 76.93 for the random effects and standard LGCP models, respectively, thus suggesting that the inclusion of $\alpha_i$ leads to improved prediction of the study counts. 

Some of the estimated effects are affected by inclusion of the random effect terms. 
For instance, the expected number of foci for verbal studies is estimated as 12.80 (95\% CI [11.57,14.14]) by the random effects LGCP as opposed to 11.67 (95\% CI [10.97,12.36]) by the fixed effects LGCP model. One possible explanation for this is that our model is assigning a low random effect to publications systematically reporting only a few foci. Such a behaviour is desired since, e.g.\, this underreporting could be solely due to author preference. Further, the random effects model provides credible intervals that fully account for the uncertainty in the regression coefficients. For example, the 95\% CI for the overall age effect $\mu_2$ provided by the fixed effects LGCP is [-0.309,-0.151], shorter than the CI provided by our model.

\section{Discussion}\label{sec:discussion}

In this work, we have presented a new CBMA model, extension of the log-Gaussian Cox process model. 
To our knowledge, this is the first application of the random effects LGCP with covariates in a 3D problem with multiple realisations. 
The model has an appealing interpretation being a spatial GLM and several interesting inferences can be obtained based on the properties of the spatial Poisson process that cannot be obtained with the commonly used kernel-based approaches. 
An advantage of our model compared to most of the existing methods is the inclusion of covariates in the analysis thus allowing for meta-regression.
Finally, a novel feature introduced in our work is the inclusion of random-effect terms which can account for additional heterogeneity in the total number of activations, compared to the standard Poisson model.

Application of our model on a meta-analysis of working memory studies have given valuable insights regarding the data. 
While our maps for the overall pattern of WM activations (Fig. \ref{fig:wm}) and the differential effect of verbal vs. non-verbal WM tasks (Fig. \ref{fig:wmdiff}) reflect previous findings found by \citet{Rottschy2012}, our fully Bayesian approach allowed us to make direct inference on probability of any foci and expected number of foci. 
Our model found no regions with  evidence of different rates of foci between verbal and non-verbal WM tasks (Appendix \ref{sec:datasupp}, Table \ref{tab:fullbrain2}). 
Importantly, our model allows a meta-regression, and we examined the effect of age and found no strong effects but generally negative effects of age on the number of foci.

There are few limitations to our work. 
Firstly, even though we found that the proposed MCMC algorithm performed well in most of the applications considered, we believe that there is room for further for improvement. 
For example, one can consider adaptive schemes in order to automatically adjust the mass matrix $\mathbf{M}$ of the HMC which we found that is crucial for the mixing properties of the algorithm. 
Secondly, we are currently not considering the problem of learning the hyperparameter $\kappa$ that controls the posterior variability of the random effect terms, but rather make use of our prior expectations to tune it. 
However, since we found that results are sensitive to the specification of $\kappa$, it is plausible to consider estimating it along with remaining model parameters.

Our work can be extended in several ways. 
One possible direction for future research is to perform a head-to-head comparison of existing methodologies that can be used for posterior inference with the proposed LGCP model in the context of CBMA. 
However, given the computation time required to apply these methods to a 3D problem, such a comparison might be too long. 
Another potential future direction is to study the conditions, such as sample size or minimum number of foci, under which it is possible to estimate several global or spatially varying effects using the LGCP. 
Such work can be of importance for practical implementations since it will provide some guidance regarding the complexity of meta-regression models that can be fit to a given dataset.

Another open problem is how to use some additional information about the foci such as $p$-values or $T$-scores. 
These values can be attached as marks to the existing point patterns. 
Such an approach can enrich the inferences obtained from a CBMA by characterising the magnitude of activation in each region as opposed to the localisation of activations, which is the question that current methods address. 
Finally, it is worth considering a zero-truncated LGCP model. 
The reason is that several CBMAs use data from databases such as BrainMap \citep{Laird2005}, where only studies with at least one focus are registered. 
For such applications, a model that does not account for the zero-truncation can provide biased intensity estimates, especially when the expected number of foci per study is low. 
Currently, very few of the existing approaches propose adjustments for this potential problem.


\bibliographystyle{bibo}
\bibliography{pbias,spatial,mcmc,cbma,neuroimaging}

\newpage
\newpage
\appendix

\section{Gradient expressions for the LGCP} \label{sec:lgcpgrad}
Let $\beta_k\equiv\mu_k$ ($k=0,\dots,K$). 
The log-posterior, up to a normalising constant is given by:
\begin{multline}
\ell\left(\boldsymbol{\alpha},\boldsymbol{\beta},\boldsymbol{\sigma},\boldsymbol{\rho},\left\{\boldsymbol{\gamma}_k\right\}_{k=1}^{K^\ast}\mid\cdot\right) 
\propto
\sum_{i=1}^{I}{
\left[-\sum_{j=1}^{V}{A_{v_j}\lambda_i(v_j)}+\sum_{j=1}^{V}{\mathbf{1}_{v_j\in\mathbf{x}_i}\log{\lambda_i(v_j)}}\right]
}
\\
+\log{\mathrm{priors}}
,\end{multline}
where $\boldsymbol\alpha=(\alpha_1,\dots,\alpha_I)^\top$, $\boldsymbol{\beta}=(\beta_0,\dots,\beta_{K})^\top$, $\boldsymbol{\sigma}=(\sigma_0,\dots,\sigma_{K^\ast})^\top$, $\boldsymbol{\rho}=(\rho_0,\dots,\rho_{K^\ast})^\top$, $A_{v_j}=\mathbf{1}_{v_j\in \mathcal{B}}$ and the intensity function at each voxel $v_j$ is defined as:
\begin{equation}
\lambda_i(v_j)=
\alpha_i
\exp{\left(\sum_{k=0}^{K}{\beta_kz_{ik}}\right)}
\exp{\left(\sum_{k=0}^{K^\ast}{\sigma_k\left(\mathbf{R}^{1/2}_k\boldsymbol\gamma_k\right)_jz_{ik}}\right)}
\end{equation}
We now calculate the derivatives with respect to the parameters of interest.
\subsection{Partial derivatives with respect to $\beta_l$}
We have that:
\begin{eqnarray}\nonumber
\frac{\partial\log{\lambda_i(v_j)}}{\partial\beta_l} &=&
\frac{\partial}{\partial\beta_l}\log{\alpha_i}+
\frac{\partial}{\partial\beta_l}\sum_{k=0}^{K}{\beta_kz_{ik}}+
\frac{\partial}{\partial\beta_l}\sum_{k=0}^{K^\ast}{\sigma_k\left(\mathbf{R}^{1/2}_k\boldsymbol\gamma_k\right)_jz_{ik}}
\\
&=&z_{il}.
\end{eqnarray}
As a result:
{
\allowdisplaybreaks
\begin{eqnarray}\nonumber
\frac{\partial\ell\left(\beta_l\mid\cdot\right)}{\partial\beta_l} & = &-\sum_{i=1}^{I}{\sum_{j=1}^{V}{\left[
A_{v_j}\frac{\partial}{\partial\beta_l}\lambda_i(v_j)-
\mathbf{1}_{v_j\in\mathbf{x}_i}\frac{\partial}{\partial\beta_l}\log{\lambda_i(v_j)}
\right]}}+
\frac{\partial}{\partial\beta_l}\log{\pi(\beta_l)} \\ \nonumber 
&=& -\sum_{i=1}^{I}{\sum_{j=1}^{V}{\left[A_{v_j}\lambda_i(v_j)z_{il}-\mathbf{1}_{v_j\in\mathbf{x}_i}z_{il}\right]}}-\frac{\partial}{\partial\beta_l}\frac{\beta_l^2}{2\tau^2}\\ 
&=&-\sum_{j=1}^{V}{\sum_{i=1}^{I}{A_{v_j}\lambda_i(v_j)z_{il}}}+\sum_{i=1}^{I}n_iz_{il}-\frac{\beta_l}{\tau^2},
\end{eqnarray}
}
where $n_i$ is the total number of foci in study $i$.

\subsection{Partial derivatives with respect to $\sigma_l$}
We have that:
\begin{eqnarray}\nonumber
\frac{\partial\log{\lambda_i(v_j)}}{\partial\sigma_l} &=&
\frac{\partial}{\partial\sigma_l}\log{\alpha_i}+
\frac{\partial}{\partial\sigma_l}\sum_{k=0}^{K}{\beta_kz_{ik}}+
\frac{\partial}{\partial\sigma_l}\sum_{k=0}^{K^\ast}{\sigma_k\left(\mathbf{R}^{1/2}_k\boldsymbol\gamma_k\right)_jz_{ik}}\\
&=&\left(\mathbf{R}^{1/2}_l\boldsymbol\gamma_l\right)_jz_{il}.
\end{eqnarray}
Therefore:
\begin{eqnarray}\nonumber
\frac{\partial\ell\left(\sigma_l\mid\cdot\right)}{\partial\sigma_l} & = &-\sum_{i=1}^{I}{\sum_{j=1}^{V}{\left[A_{v_j}\frac{\partial}{\partial\sigma_l}\lambda_i(v_j)-\mathbf{1}_{v_j\in\mathbf{x}_i}\frac{\partial}{\partial\sigma_l}\log{\lambda_i(v_j)}\right]}}+\frac{\partial}{\partial\sigma_l}\log{\pi(\sigma_l)} \\ \nonumber 
&=& -\sum_{i=1}^{I}{\sum_{j=1}^{V}{\left[A_{_vj}\lambda_i(v_j)\left(\mathbf{R}^{1/2}_l\boldsymbol\gamma_l\right)_jz_{il}-\mathbf{1}_{v_j\in\mathbf{x}_i}\left(\mathbf{R}^{1/2}_l\boldsymbol\gamma_l\right)_jz_{il}\right]}}-\frac{\partial}{\partial\sigma_l}\frac{\sigma_l^2}{2\tau^2}\\ \nonumber
&=&-\sum_{j=1}^{V}{\sum_{i=1}^{I}{\left[\left(\mathbf{R}^{1/2}_l\boldsymbol\gamma_l\right)_j\left(A_{v_j}\lambda_i(v_j)z_{il}-\mathbf{1}_{v_j\in\mathbf{x}_i}z_{il}\right)\right]}}-\frac{\sigma_l}{\tau^2}\\ \nonumber
&=&-\sum_{j=1}^{V}{\left[\left(\mathbf{R}^{1/2}_l\boldsymbol\gamma_l\right)_j\sum_{i=1}^{I}{\left[A_{v_j}\lambda_i(v_j)z_{il}-\mathbf{1}_{v_j\in\mathbf{x}_i}z_{il}\right]}\right]}-\frac{\sigma_l}{\tau^2}.
\end{eqnarray}

\subsection{Partial derivatives with respect to $\rho_l$}\label{subsec:appenrho}
Again:
\begin{eqnarray}\label{appenrhoa}\nonumber
\frac{\partial\log{\lambda_i(v_j)}}{\partial\rho_l} &=&
\frac{\partial}{\partial\rho_l}\log{\alpha_i}+
\frac{\partial}{\partial\rho_l}\sum_{k=0}^{K^\ast}{\beta_kz_{ik}}+
\frac{\partial}{\partial\rho_l}\sum_{k=0}^{K}{\sigma_k\left(\mathbf{R}^{1/2}_k\boldsymbol\gamma_k\right)_jz_{ik}}\\
&=&\sigma_l\frac{\partial}{\partial\rho_l}\left(\mathbf{R}^{1/2}_l\boldsymbol\gamma_l\right)_jz_{il}.
\end{eqnarray}
For ease of exposition we complete the derivation for the one-dimensional case; however, similar arguments can be used when $\mathcal{B}\subset\mathbb{R}^3$. 
Matrices $\mathbf{R}_l$ are circulant and so, the matrix-vector product $\mathbf{R}^{1/2}_l\boldsymbol\gamma_l$ can be found using the discrete Fourier transform as
\begin{equation}\label{appenrhob}
\mathbf{R}^{1/2}_l\boldsymbol\gamma_l = \mathbf{F}\boldsymbol\Phi_l^{1/2}\mathbf{F}^\mathrm{H}\boldsymbol\gamma_l,
\end{equation}
where $\boldsymbol\Phi_l$ are the diagonal matrices containing the eigenvalues of $\mathbf{R}_l$ and $\mathbf{F}$ is the matrix of eigenvectors. In Equation \eqref{appenrhob}, the only term depending on $\rho_l$ is $\boldsymbol\Phi_l$ and, hence:
\begin{equation}
\frac{\partial}{\partial\rho_l}\mathbf{R}^{1/2}_l\boldsymbol\gamma_l = \mathbf{F}\frac{\partial}{\partial\rho_l}\boldsymbol\Phi_l^{1/2}\mathbf{F}^\mathrm{H}\boldsymbol\gamma_l
\end{equation}
We know that $\boldsymbol\Phi_l=\mathrm{diag}\left\{\phi_{l_0},\dots,\phi_{l_{V-1}}\right\}$, where for $k=0,\dots,V-1$ we have that:
\begin{equation}\label{appenrhoc}
\phi_{l_k}=\sum_{j=0}^{V-1}{\exp{\left(-\rho_l ||v_0,v_j||^{\delta_l}\right)}\exp{\left(-\frac{2\pi\iota kj}{V}\right)}},
\end{equation}
where $\iota$ is the imaginary unit. Now it is straightforward to see that for $k=0,\dots,V-1$:
\begin{eqnarray}\nonumber
\frac{\partial}{\partial\rho_l}\phi_{l_k}^{1/2}&=&\frac{\partial}{\partial\rho_l}\sqrt{\sum_{j=0}^{V-1}{\exp{\left(-\rho_l ||v_0,v_j||^{\delta_l}\right)}\exp{\left(-\frac{2\pi\iota kj}{V}\right)}}}\\ \nonumber
&=&\frac{\frac{\partial}{\partial\rho_l}\sum_{j=0}^{V-1}{\exp{\left(-\rho_l ||v_0,v_j||^{\delta_l}\right)}\exp{\left(-\frac{2\pi\iota kj}{V}\right)}}}{2\sqrt{\sum_{j=0}^{V-1}{\exp{\left(-\rho_l ||v_0,v_j||^{\delta_l}\right)}\exp{\left(-\frac{2\pi\iota kj}{V}\right)}}}}\\ \nonumber
&=&\frac{-\sum_{j=1}^{V-1}{d\left(v_0,v_j\right)^{\delta_l}\exp{\left(-\rho_l ||v_0,v_j||^{\delta_l}\right)}\exp{\left(-\frac{2\pi\iota kj}{V}\right)}}}{2\phi_{l_k}^{1/2}}\\
&=&-\frac{1}{2}\frac{\psi_{l_k}}{\phi_{l_k}^{1/2}},
\end{eqnarray}
where $\psi_{l_k}$ can be viewed as the $k$-th eigenvalue of the of a circulant matrix $\mathbf{S}_l$ with base $\mathbf{s}_l=\left[||v_0,v_0||^{\delta_l}\exp{\left(-\rho_l ||v_0,v_0||^{\delta_l}\right)},\dots, ||v_0,v_{V-1}||^{\delta_l}\exp{\left(-\rho_l||v_0,v_{V-1}||^{\delta_l}\right)}\right]$ and $\mathbf{S}_l=\mathbf{F}\boldsymbol\Psi_l\mathbf{F}^{\mathrm{H}}$, $\boldsymbol\Psi_l=\mathrm{diag}\left\{\psi_{l_0},\dots,\psi_{l_{V-1}}\right\}$. Overall we see that:
\begin{eqnarray}\label{appenrhoc}\nonumber
\frac{\partial}{\partial\rho_l}\mathbf{R}^{1/2}_l\boldsymbol\gamma_l &=& \mathbf{F}\frac{\partial}{\partial\rho_l}\boldsymbol\Phi_l^{1/2}\mathbf{F}^\mathrm{H}\boldsymbol\gamma_l\\ \nonumber
&=&-\frac{1}{2}\mathbf{F}\left[\boldsymbol\Psi_l\oslash\boldsymbol\Phi^{1/2}\right]\mathbf{F}^{\mathrm{H}}\boldsymbol\gamma_l\\
&=&-\frac{1}{2}\mathbf{Q}_l\boldsymbol\gamma_l,
\end{eqnarray}
where $\oslash$ stands for element wise division. Combining Equations \eqref{appenrhoa} and \eqref{appenrhoc}, we find that:
\begin{equation}
\frac{\partial\log{\lambda_i(v_j)}}{\partial\rho_l} = -\frac{1}{2}\sigma_l\left(\mathbf{Q}_l\boldsymbol\gamma_l\right)_jz_{il}.
\end{equation}
So:
\begin{eqnarray}\nonumber
\frac{\partial\ell\left(\rho_l\mid\cdot\right)}{\partial\rho_l} & = &-\sum_{i=1}^{I}{\sum_{j=1}^{V}{\left[A_{v_j}\frac{\partial}{\partial\rho_l}\lambda_i(v_j)-\mathbf{1}_{v_j\in\mathbf{x}_i}\frac{\partial}{\partial\rho_l}\log{\lambda_i(v_j)}\right]}}+\frac{\partial}{\partial\rho_l}\log{\pi(\rho_l)} \\ \nonumber 
&=& -\sum_{i=1}^{I}{\sum_{j=1}^{V}{\left[A_{_vj}\lambda_i(v_j)\left(-\frac{1}{2}\right)\sigma_l\left(\mathbf{Q}_l\boldsymbol\gamma_l\right)_jz_{il}-\mathbf{1}_{v_j\in\mathbf{x}_i}\left(-\frac{1}{2}\right)\sigma_l\left(\mathbf{Q}_l\boldsymbol\gamma_l\right)_jz_{il}\right]}}
\\ \nonumber
&&-\frac{\partial}{\partial\rho_l}\mathbf{1}_{\rho_l\in[\rho_\mathrm{low},\rho_\mathrm{upp}]}
\\ \nonumber
&=&\frac{\sigma_l}{2}\sum_{j=1}^{V}{\sum_{i=1}^{I}{\left[\left(\mathbf{Q}_l\boldsymbol\gamma_l\right)_j\left(A_{v_j}\lambda_i(v_j)z_{il}-\mathbf{1}_{v_j\in\mathbf{x}_i}z_{il}\right)\right]}}
\\ \nonumber
&=&\frac{\sigma_l}{2}\sum_{j=1}^{V}{\left[\left(\mathbf{Q}_l\boldsymbol\gamma_l\right)_j\sum_{i=1}^{I}{\left[A_{v_j}\lambda_i(v_j)z_{il}-\mathbf{1}_{v_j\in\mathbf{x}_i}z_{il}\right]}\right]}.
\end{eqnarray}

\subsection{Partial derivatives with respect to $\boldsymbol\gamma_l$}
Finally:
{
\allowdisplaybreaks
\begin{eqnarray}\nonumber
\frac{\partial\log{\lambda_i(v_j)}}{\partial\boldsymbol\gamma_l} &=&
\frac{\partial}{\partial\boldsymbol\gamma_l}\log{\alpha_i}+
\frac{\partial}{\partial\boldsymbol\gamma_l}\sum_{k=0}^{K^\ast}{\beta_kz_{ik}}+
\frac{\partial}{\partial\boldsymbol\gamma_l}\sum_{k=0}^{K}{\sigma_k\left(\mathbf{R}^{1/2}_k\boldsymbol\gamma_k\right)_jz_{ik}}\\
&=&\sigma_l\mathbf{r}_{l_j}z_{il},
\end{eqnarray}
where $\mathbf{r}_{l_j}$ is the $j$-th row of the matrix $\mathbf{R}_l^{1/2}$. Now we can see that:
\begin{eqnarray}\nonumber
\frac{\partial\ell\left(\boldsymbol\gamma_l\mid\cdot\right)}{\partial\boldsymbol\gamma_l} & = &-\sum_{i=1}^{I}{\sum_{j=1}^{V}{\left[A_{v_j}\frac{\partial}{\partial\boldsymbol\gamma_l}\lambda_i(v_j)-\mathbf{1}_{v_j\in\mathbf{x}_i}\frac{\partial}{\partial\boldsymbol\gamma_l}\log{\lambda_i(v_j)}\right]}}+\frac{\partial}{\partial\boldsymbol\gamma_l}\log{\pi(\boldsymbol\gamma_l)} \\ \nonumber 
&=& -\sum_{i=1}^{I}{\sum_{j=1}^{V}{\left[A_{v_j}\lambda_i(v_j)\sigma_l\mathbf{r}_{l_j}z_{il}-\mathbf{1}_{v_j\in\mathbf{x}_i}\sigma_l\mathbf{r}_{l_j}z_{il}\right]}}-\frac{\partial}{\partial\boldsymbol\gamma_l}\frac{\boldsymbol\gamma_l^{\mathrm{T}}\boldsymbol\gamma_l}{2}\\ \nonumber
&=&-\sigma_l\sum_{j=1}^{V}{\left[\mathbf{r}_{l_j}\sum_{i=1}^{I}{\left[A_{v_j}\lambda_i(v_j)z_{il}-\mathbf{1}_{v_j\in\mathbf{x}_i}z_{il}\right]}\right]}-\boldsymbol\gamma_l\\ \nonumber
&=&-\sigma_l\sum_{j=1}^{V}{\left[\mathbf{r}_{l_j}\mathbf{c}_{l_j}\right]}-\boldsymbol\gamma_l\\ \nonumber
&=&-\sigma_l\left(\mathbf{R}_{l}^{1/2}\right)^{\top}\mathbf{c}_l-\boldsymbol\gamma_l\\
&=&-\sigma_l\mathbf{R}_{l}^{1/2}\mathbf{c}_l-\boldsymbol\gamma_l,
\end{eqnarray}
}
since $\mathbf{R}$ is a nested block circulant matrix, where $\mathbf{c}_l$ are $V$-vectors with elements $\mathbf{c}_{l_j}=\sum_{i=1}^{I}{\left[A_{v_j}\lambda_i(v_j)z_{il}- \mathbf{1}_{v_j\in\mathbf{x_i}}z_{il}\right]}$.

\subsection{Random effects updates}\label{sec:rfxgibbs}
Let $J$ be the total number of publications from which the $I$ studies in the meta-analysis have been retrieved, and let $\mathcal{C}_j$ the set of studies retrieved from paper $j$ ($j=1,\dots,J$). 
For all $j=1,\dots,J$ we have that
\begin{eqnarray}\nonumber
\pi(\alpha_j|\cdot) & \propto & \pi(\alpha_j) \prod_{i\in\mathcal{C}_j}\pi(\mathbf{x}_i|\lambda_i)
 \\ \nonumber
 & \propto & 
 \alpha_j^{\kappa-1}\exp{\left\{-\alpha_j\kappa\right\}}
 \exp{\left\{-\alpha_j\sum_{i\in\mathcal{C}_j}\sum_{v=1}{\lambda_{iv}^\ast} \right\}}\alpha_j^{\sum_{i\in\mathcal{C}_j}n_i}
  \\ \nonumber
  & = & \alpha_j^{\sum_{i\in\mathcal{C}_j}n_i+\kappa-1}\exp{\left\{-\alpha_j\left(\sum_{i\in\mathcal{C}_j}\sum_{v=1}{\lambda_{iv}^\ast}+\kappa\right) \right\}}.
\end{eqnarray}
Therefore, we draw $\alpha_j$ from a $\mathcal{G}\left(\sum_{i\in\mathcal{C}_j}n_i+\kappa,\sum_{i\in\mathcal{C}_j}\sum_{v=1}{\lambda_{iv}^\ast}+\kappa\right)$ distribution.

\newpage
\section{Simulation studies}\label{sec:simulations}
In order evaluate the performance of the proposed HMC algorithm to sample from the posterior distribution of the latent GPs, we consider two simulation setups. 
In the first we draw samples directly from the log-Gaussian Cox process model, whereas in the second we create synthetic studies based on a different model to assess its robustness to model misspecification. 
For consistency, all processesare defined on the same brain atlas used in the application of Section \ref{sec:data}, consisting of 216,040 $2\mathrm{mm}^3$ cubic voxels. 
The average number of foci per simulated dataset is kept low (mean number of foci per study is 5) to resemble the sparsity of points observed in real CBMA data. 
Finally, the total number of studies is fixed to 200 in both analyses, similar to the sample sizes available in real applications \citep[for example]{Kang2011}.

\subsection{Setup 1}
\label{subsec:simI}
In this setting we simulate 200 studies, with two spatially varying covariates that account for the mean of two groups of studies, and two non-spatially varying covariates. 
For $i=1,\dots,200$ we set:
\begin{equation}
\lambda_{iv} = \exp{\left\{ \sum_{k=1}^{2}{\left(\mu_k+\sigma_k\left(\mathbf{R}^{1/2}_k\boldsymbol\gamma_k\right)_v\right)z_{ik}} +\sum_{i=3}^{4}{\beta_kz_{ik}} \right\}},
\end{equation}
where $z_{i1}\sim\mathrm{Bernoulli}(0.5)$, $z_{i2}=1-z_{i1}$, $z_{i3}\sim\mathrm{Uniform}[-1,1]$ and $z_{i4}\sim\mathrm{Bernoulli}(0.5)$. 
Note that this parametrisation of the covariates implies existence of two types of studies, say type 1 and 2, with different spatially varying means and the effect of one continuous and one categorical covariate. 
The expected total number of foci is 3.99 and 4.16 for studies of type 1 and 2 respectively. 
We draw $\boldsymbol\gamma_1, \boldsymbol\gamma_2$ from their  $\mathcal{N}_V(0,\mathbf{I})$ prior  and fix the values of the scalar parameters shown in Table \ref{tab:lgcpsim1}.   
We run the HMC algorithm of Section \ref{sec:algorithm} for 10,000 iterations, discarding the first 4,000 as a burn-in and save every 6 iterations for a total of 1,000 saved posterior samples.  
This took roughly 14 hours on an NVIDIA Tesla K20c GPU card.  

\begin{figure}[htp]
\centering
\includegraphics[scale=0.31]{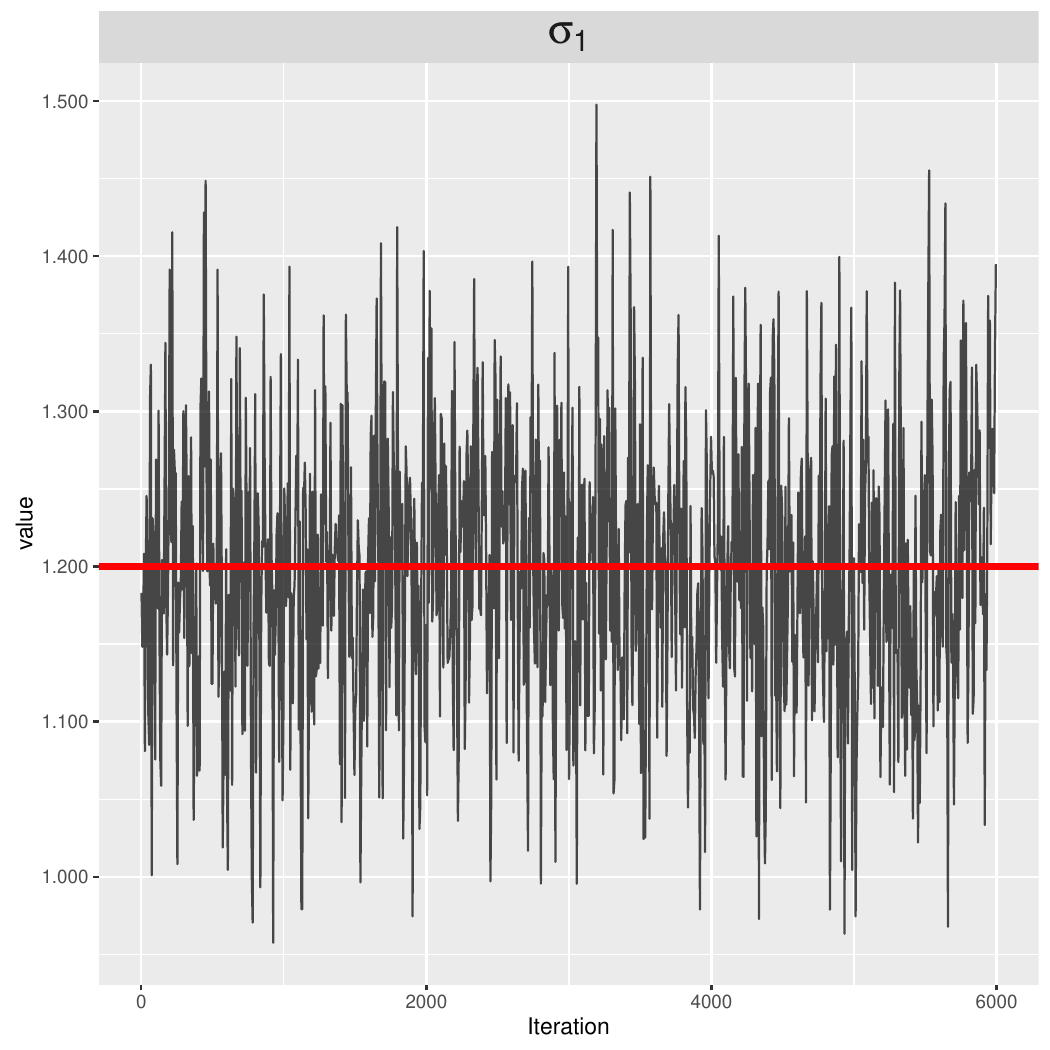}
\includegraphics[scale=0.31]{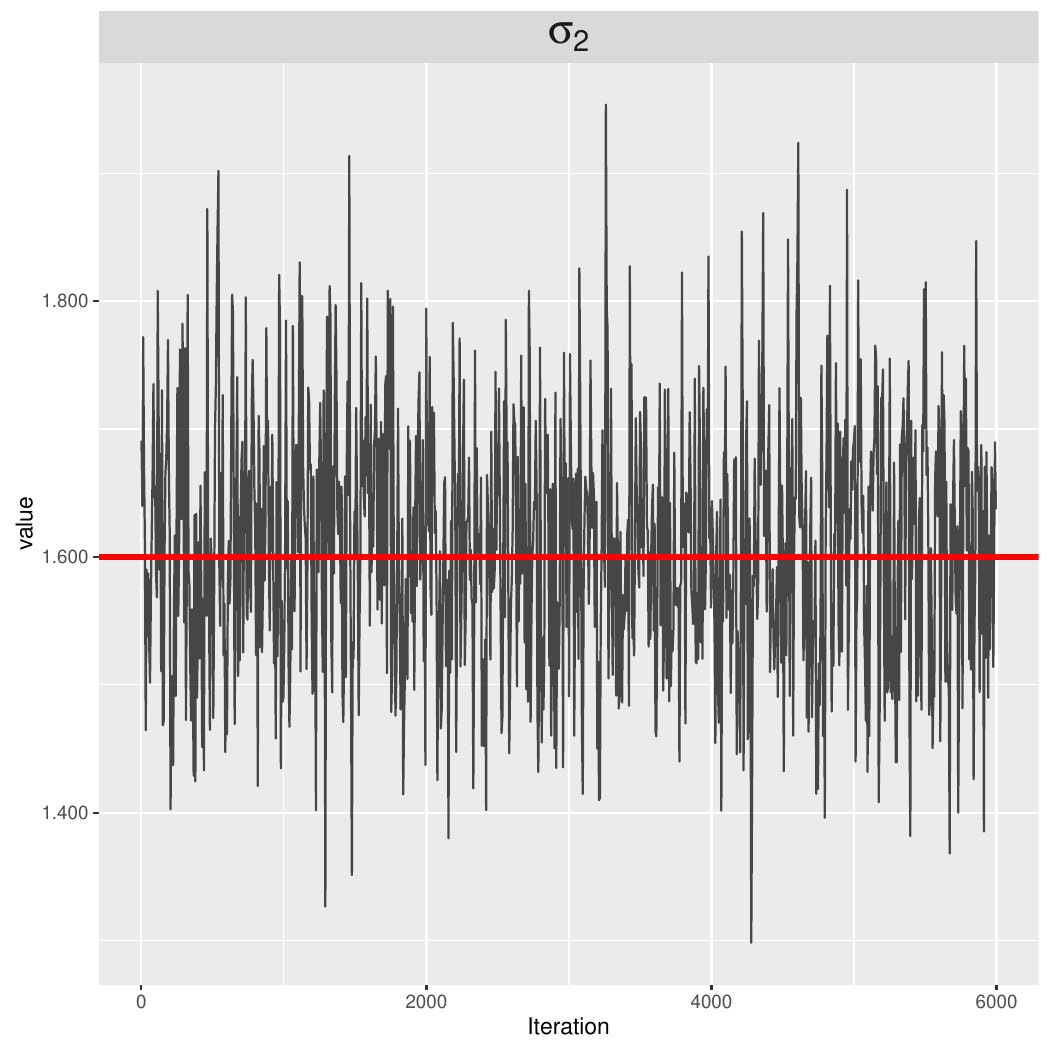} \\
\includegraphics[scale=0.31]{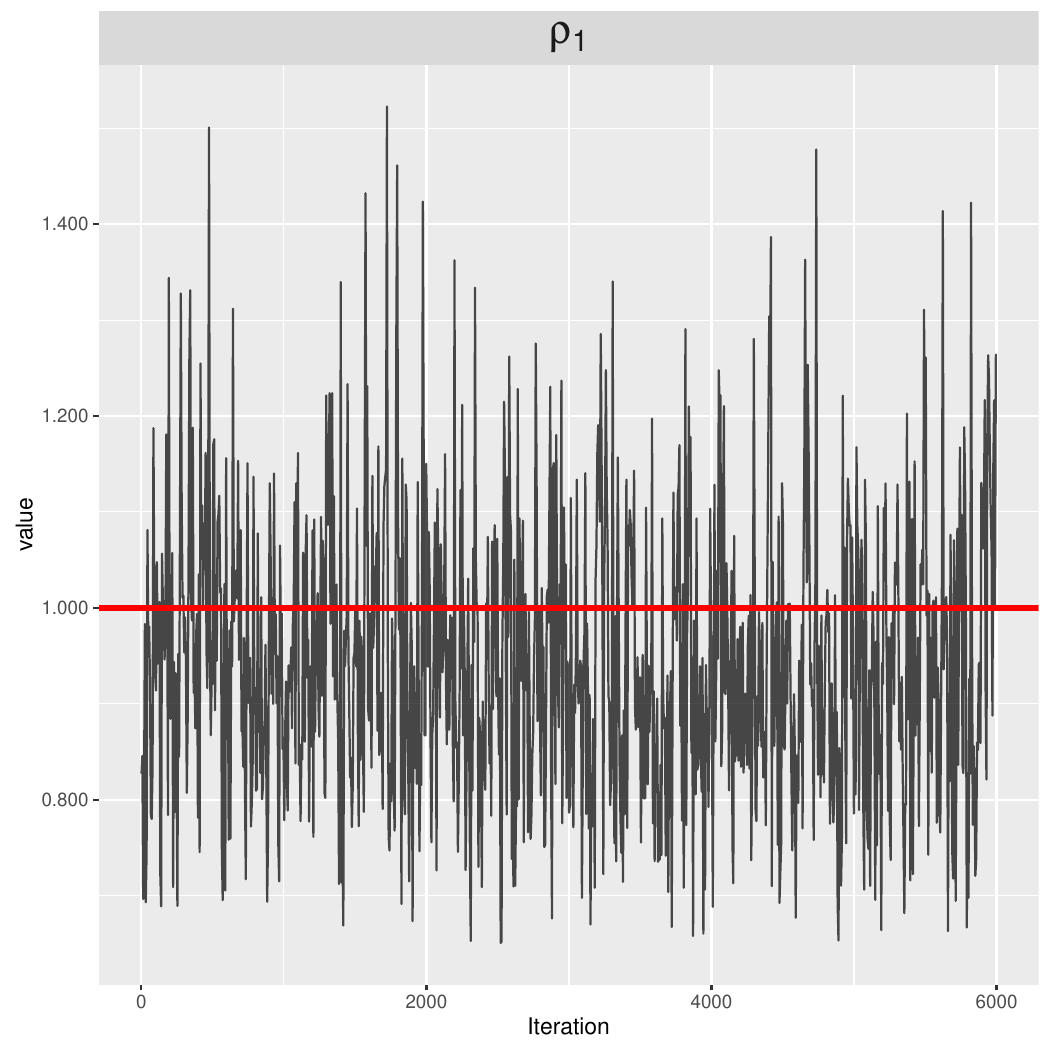} 
\includegraphics[scale=0.31]{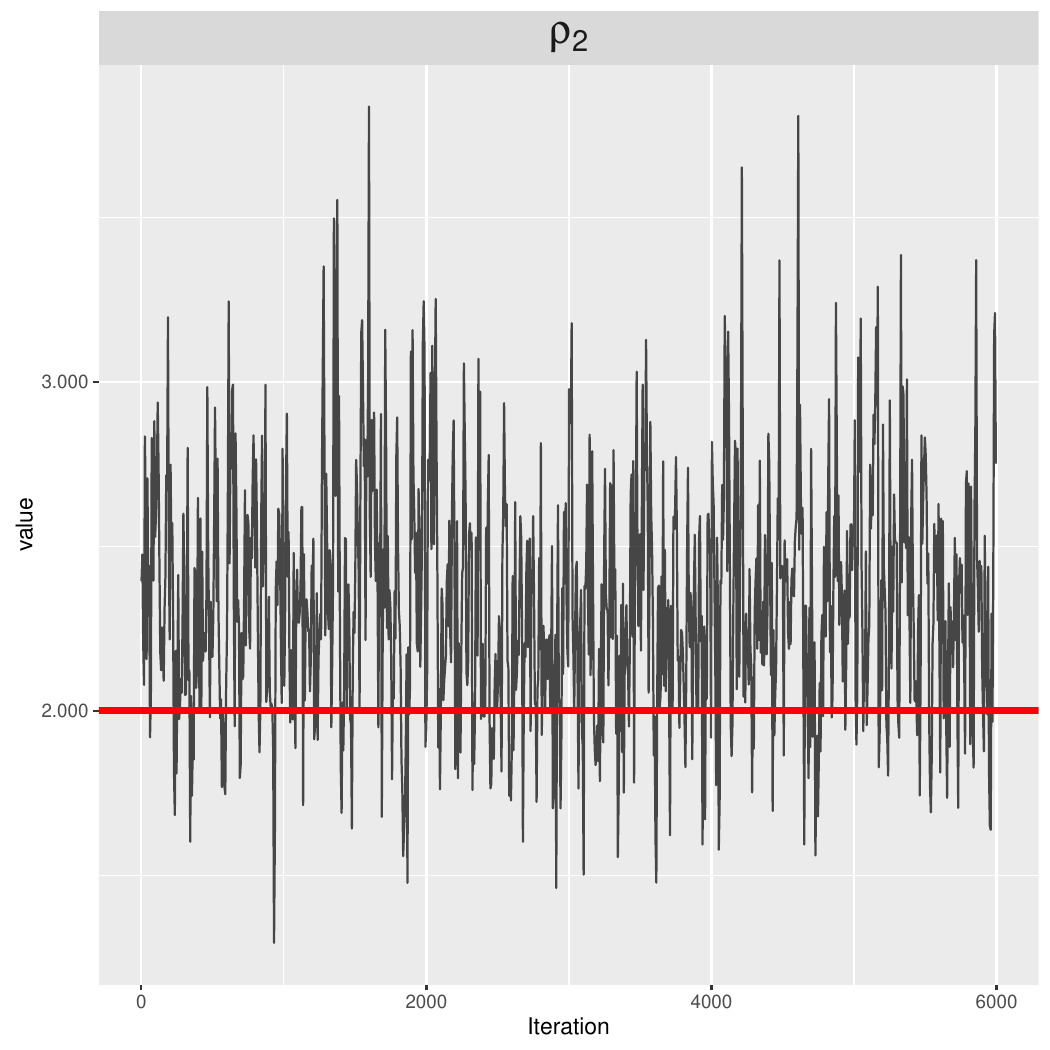} 
\caption{Posterior traceplots for the scalar parameters of the LGCP model used to fit the data of  Section \ref{subsec:simI}. Top row: standard deviations. Bottom row: correlation decay parameters ($\times 100$). The true values are indicated by the solid red lines.}
\label{fig:lgcpsim3}
\end{figure}

\begin{figure}[htp]
\centering
\includegraphics[scale=0.31]{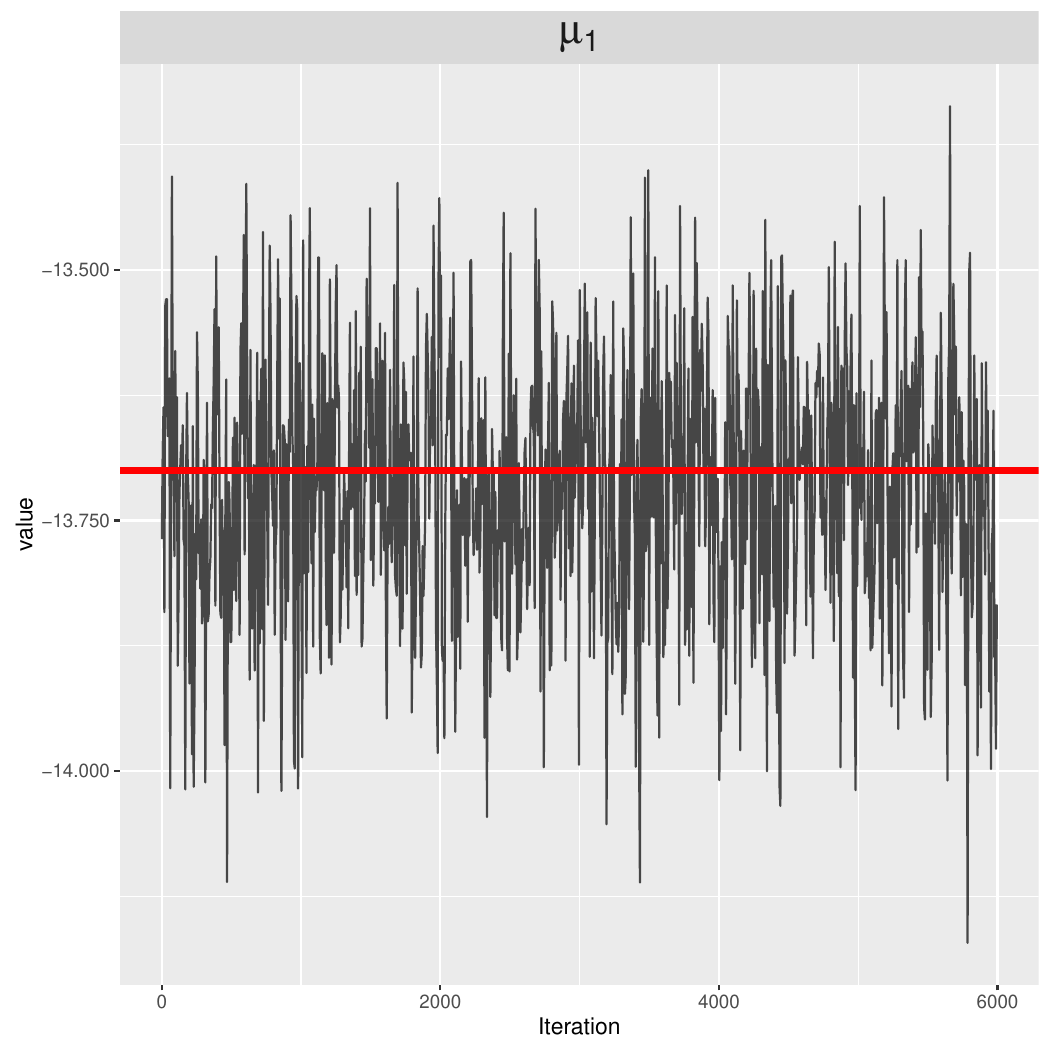}
\includegraphics[scale=0.31]{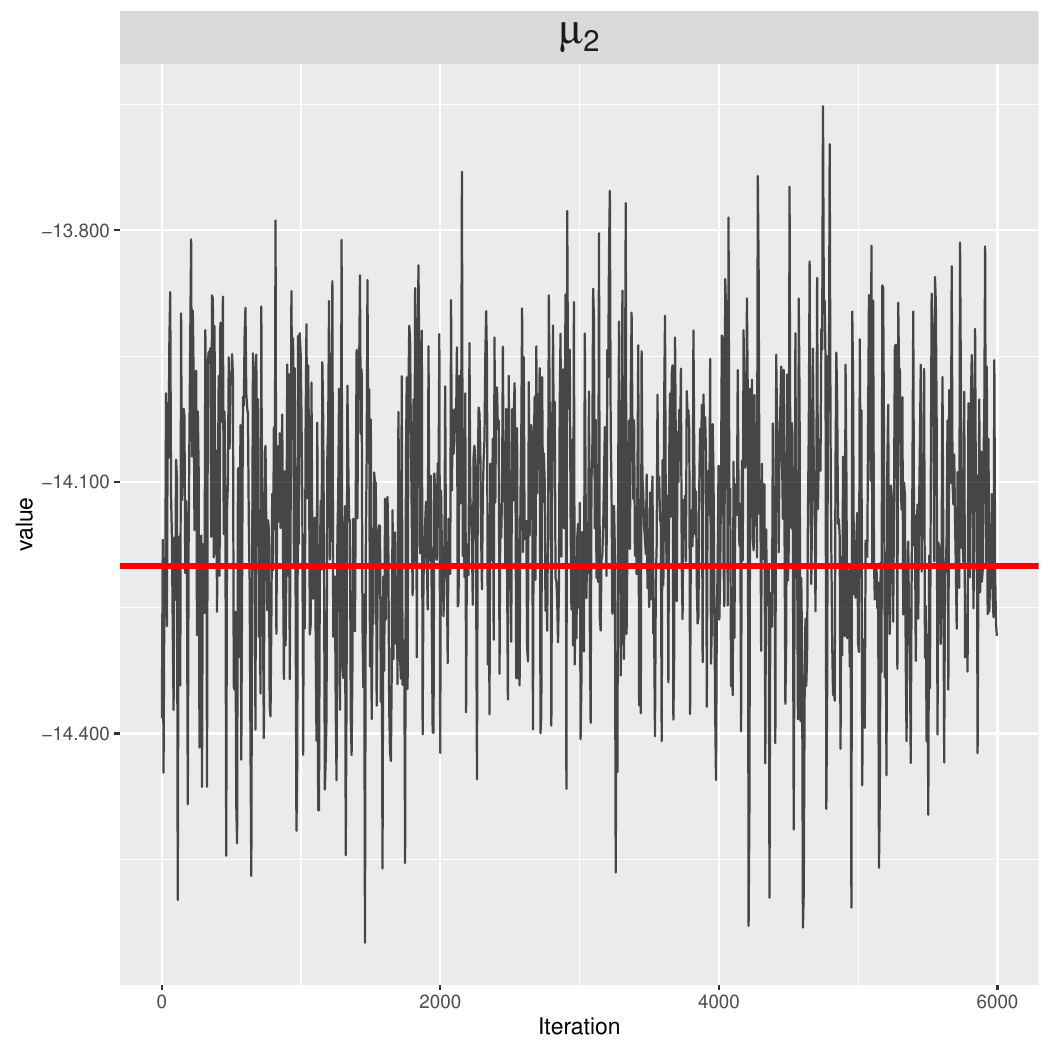} \\
\vspace{-0.2in}
\includegraphics[scale=0.31]{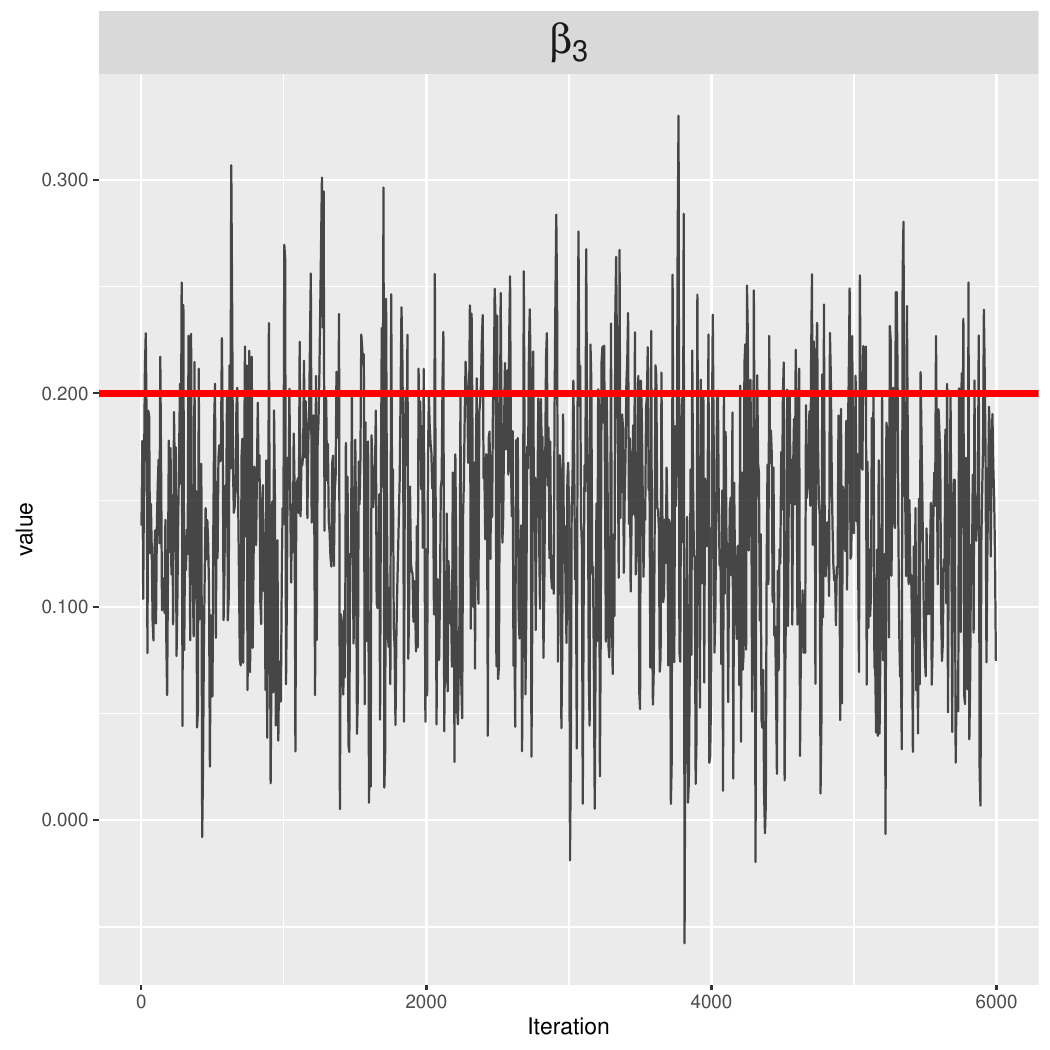} 
\includegraphics[scale=0.31]{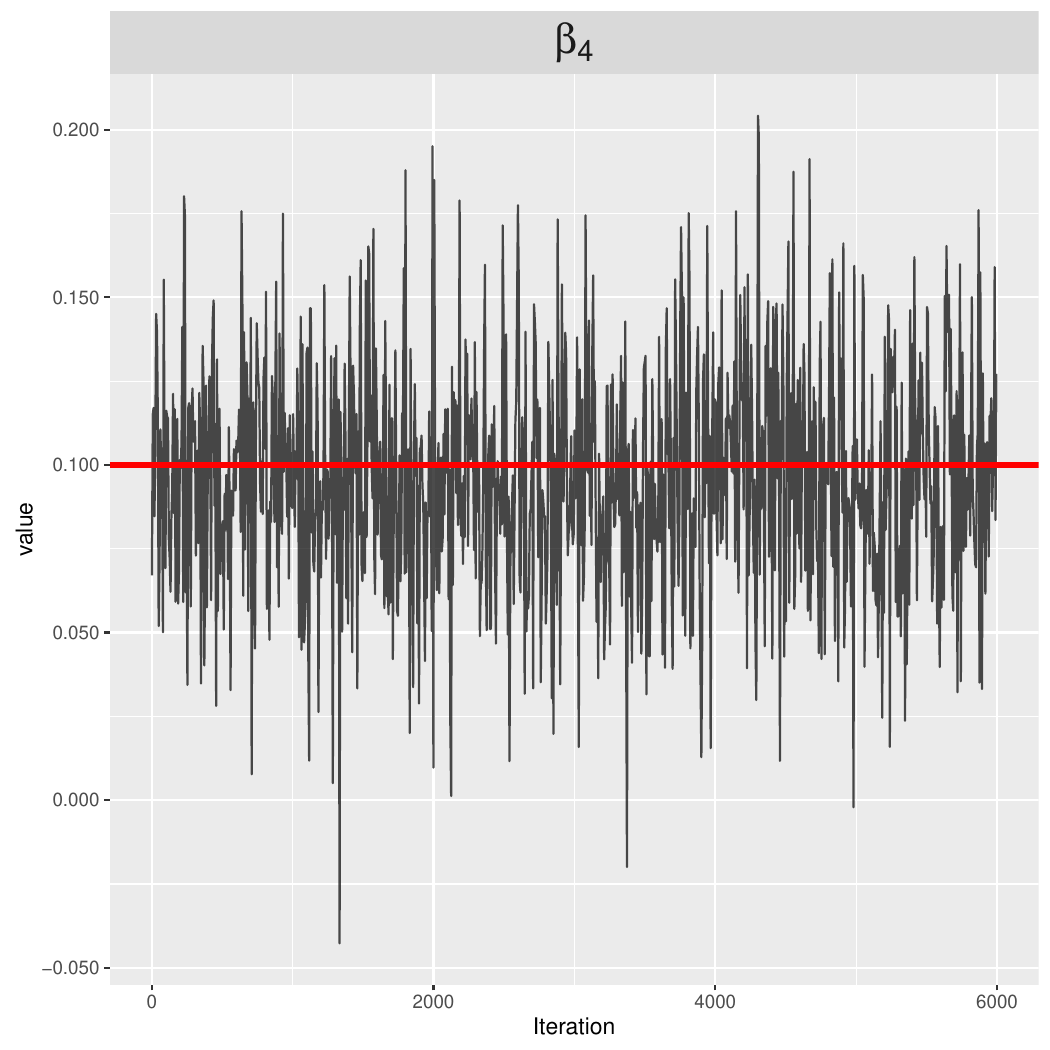} 
\caption[LGCP simulation study 1: traceplots for $\mu_k$ and $\beta_k$]{Posterior traceplots for the scalar parameters of the LGCP model used to fit the data of  Section \ref{subsec:simI}. Top row: overall latent process means. Bottom row: regression coefficients for covariates $z_3$ and $z_4$. The true values are indicated by the solid red lines.}
\label{fig:lgcpsim4}
\end{figure}

\begin{landscape}
\begin{figure}[htp]
  \centering
  \vspace{-1.5in}
\includegraphics[scale=1.05]{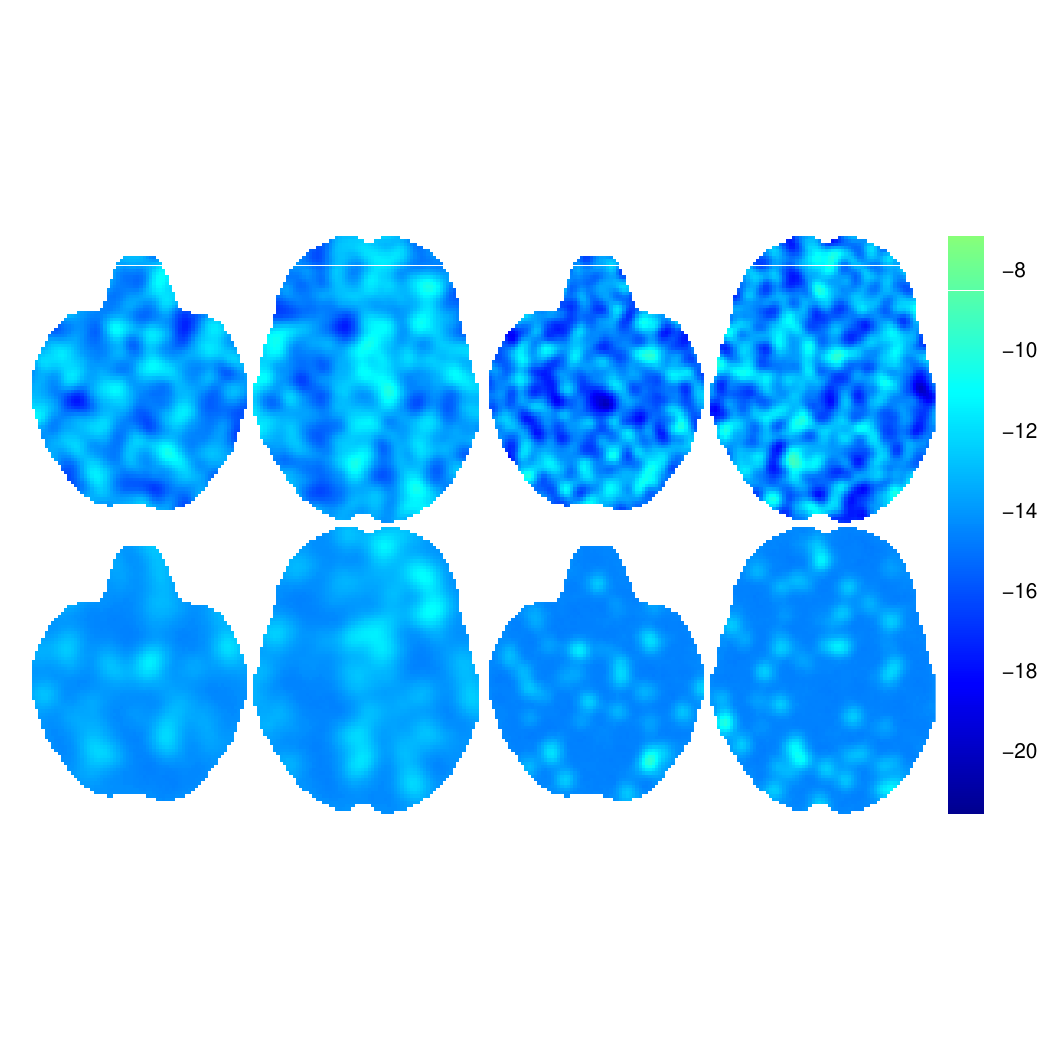}
\vspace{-1.6in}
\caption{Some true (top row) and estimated (bottom row) latent Gaussian processes for type 1 (columns 1 and 2) and type 2 (columns 3 and 4) in the simulation setup 1 of Section \ref{subsec:simI}. Columns 1 and 3 correspond to axial slice $z=-22$; columns 2 and 4 correspond to axial slice $z=4$. While they may appear dissimilar at first, observe that the most intense regions of the true and estimated intensity match up; in this punishingly sparse setting (mean $\approx$ 4 foci per 3D image), the less intense regions have too few points to learn the intensity.}
  \label{fig:sim1comb}
 \end{figure}
\end{landscape}

Results are summarised in Table \ref{tab:lgcpsim1} and Figure \ref{fig:sim1comb}. 
In Table \ref{tab:lgcpsim1} we see that the scalar parameters are estimated accurately despite the sparsity of points in the realisations.
The 95\% credible intervals contain the true values of all the parameters in the setup. 
Traceplots for the parameters $\sigma_1$, $\sigma_2$, $\rho_1$ and $\rho_2$ can be found in Figure \ref{fig:lgcpsim3}, whereas trace plots for $\mu_1$, $\mu_2$, $\beta_3$ and $\beta_4$ can be found in Figure \ref{fig:lgcpsim4}. 
The red lines indicate the true parameter values.  

For $z_{i3}=z_{i4}=0$, the median expected number of points is 3.97 (95\% CI [3.84,4.10])for type 1 and 4.61 (95\% CI [4.46,4.78]) for type 2. 
These values are very similar to the values we observe in the simulated dataset, that is 3.98 for type 1 and 4.53  for type 2. 
This indicates that our model does a good job fitting the data.  
The shape of the latent Gaussian processes $\mu_k+\sigma_k\mathbf{R}^{1/2}_k\boldsymbol\gamma_k$ is generally captured for both types as can be seen in Figure \ref{fig:sim1comb}. 
In particular, we can see that the maxima in the true and estimated images appear roughly in the same locations. 
The same cannot be said about the other values but this is expected given the dearth of information in regions of low intensity.

\begin{longtable}{cccc}
\caption{Posterior summaries of the scalar parameters of the LGCP model, fit to the simulated data of Section \ref{subsec:simI}. Results are based on 1,000 posterior draws. The values for the correlation parameters $\rho_1$, $\rho_2$ are multiplied by 100. The values for $\beta_1$ and $\beta_2$ are multiplied by 10.} 
\label{tab:lgcpsim1}\\
\hline
\textbf{Parameter} & \textbf{True Value}& \textbf{Posterior median} & \textbf{$95\%$ credible interval} \\ \hline 
$\mu_1$&-13.7&-13.72 & -13.99 , -13.48 \\
$\mu_2$&-14.2&-14.14 & -14.47 , -13.86 \\
$\sigma_1$&1.2&1.19 & 1.01 , 1.38 \\
$\sigma_2$&1.6&1.61 & 1.43 , 1.81  \\
$\rho_1$&1&0.93& 0.69 , 1.27 \\
$\rho_2$&2&2.30& 1.69 , 3.15  \\
$\beta_3$&2&1.44 & 0.22 , 2.52 \\
$\beta_4$&1&0.95 & 0.32 , 1.65 \\
\hline
\end{longtable}

\subsection{Setup 2}
\label{subsec:simII}
\begin{landscape}
\begin{figure}[htp]
        \centering
        \hspace{+1.5in}
        \vspace{-0.2in}
        \includegraphics[scale=0.4]{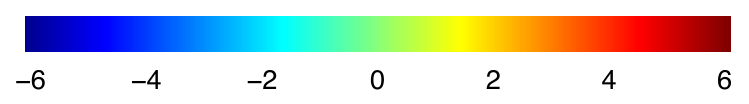}
        \\
        \includegraphics[scale=0.30]{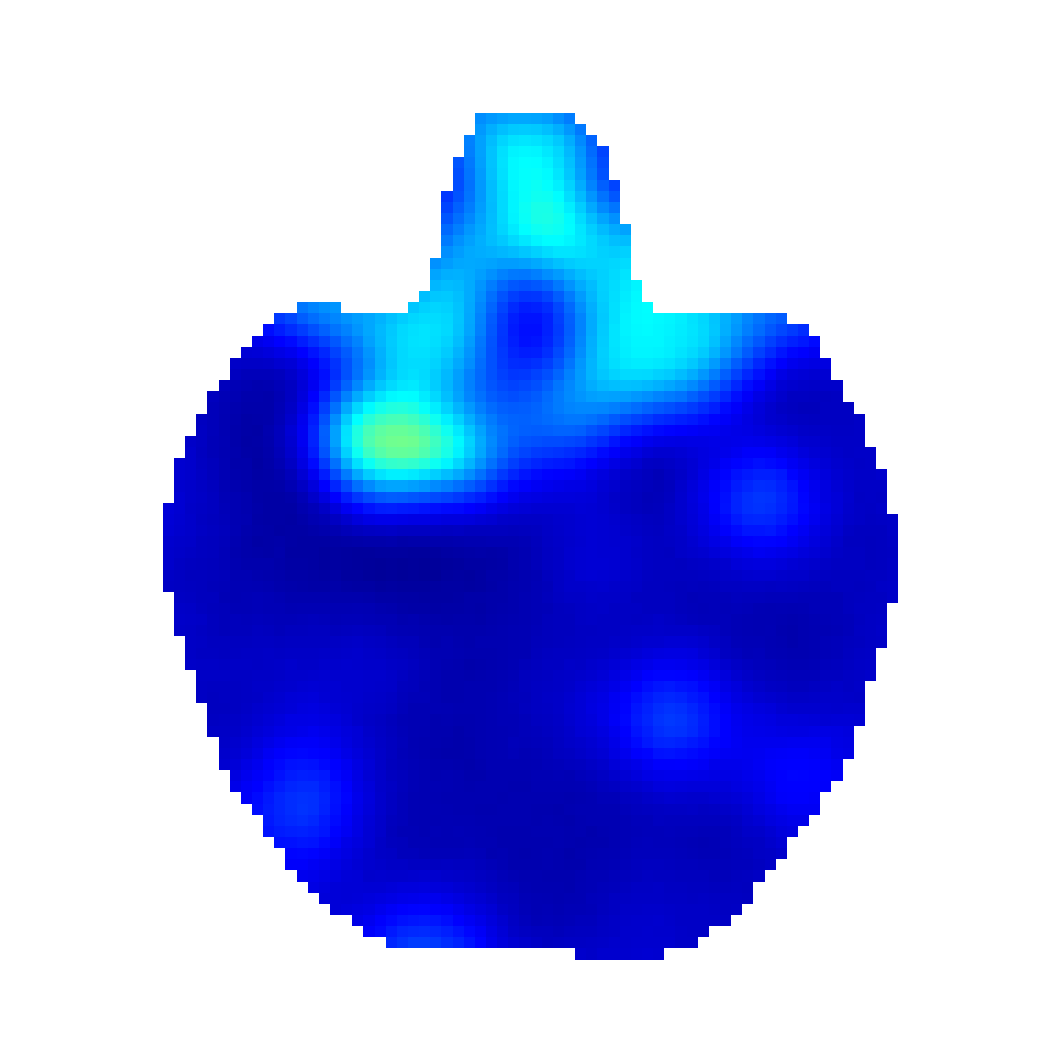}
        \hspace{-0.65in}
        \includegraphics[scale=0.30]{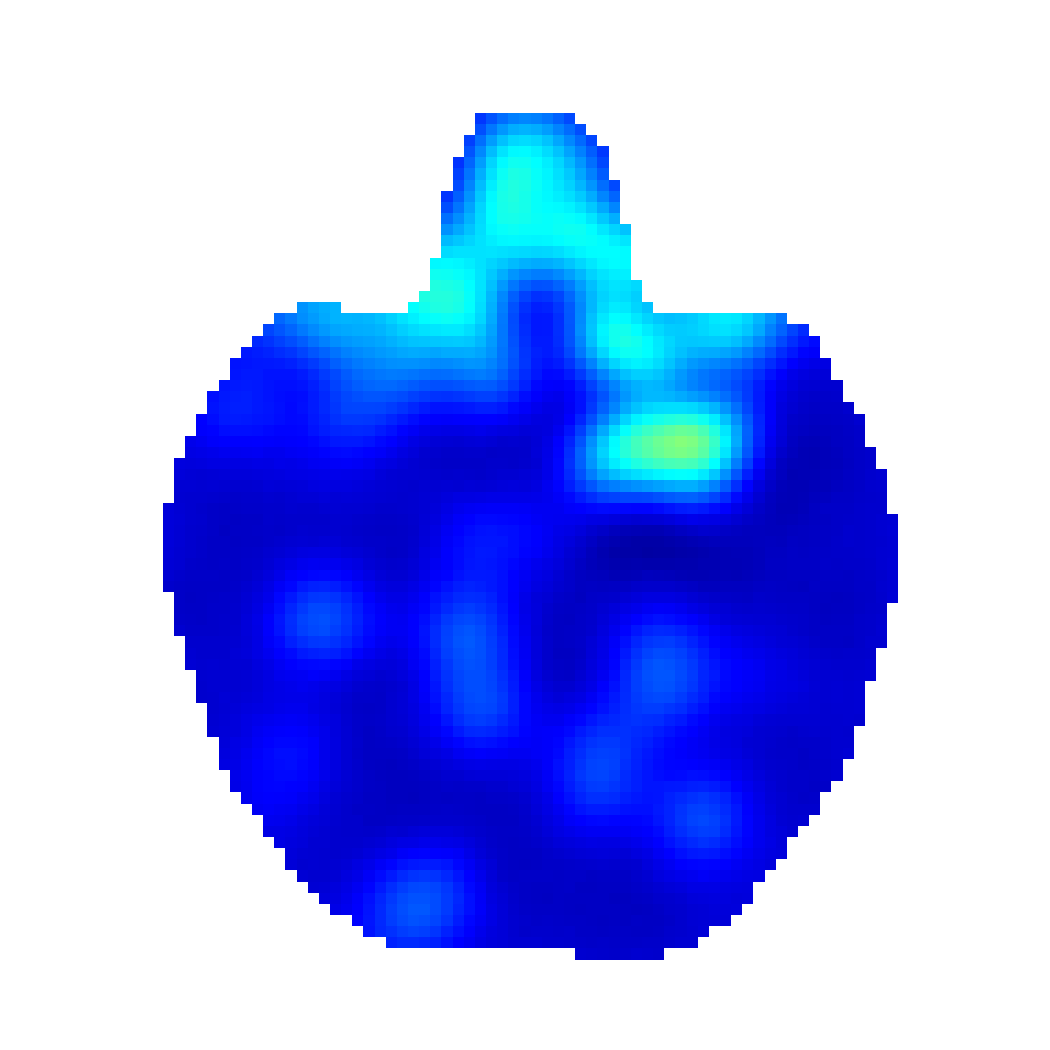}
         \hspace{-0.65in}
        \includegraphics[scale=0.30]{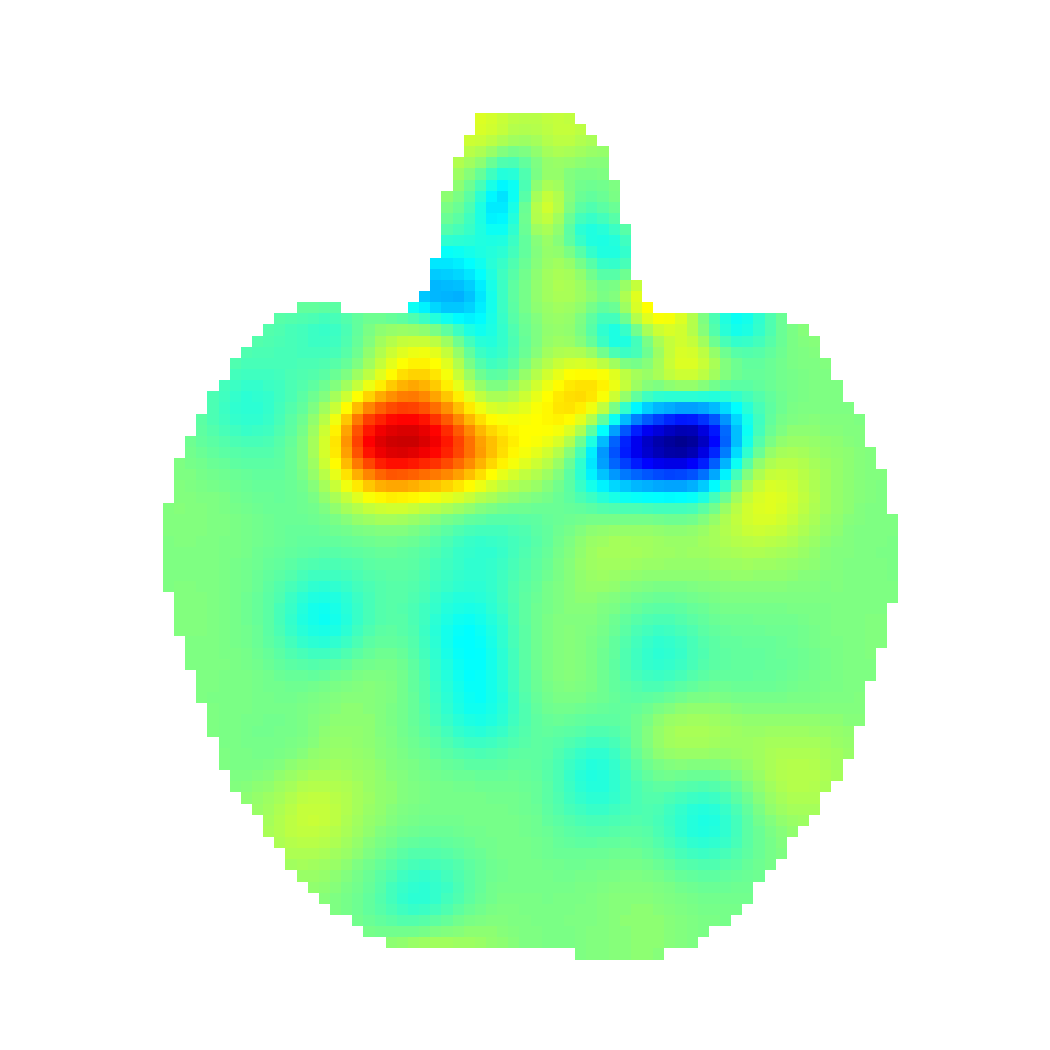}
         \hspace{-0.65in}
        \includegraphics[scale=0.30]{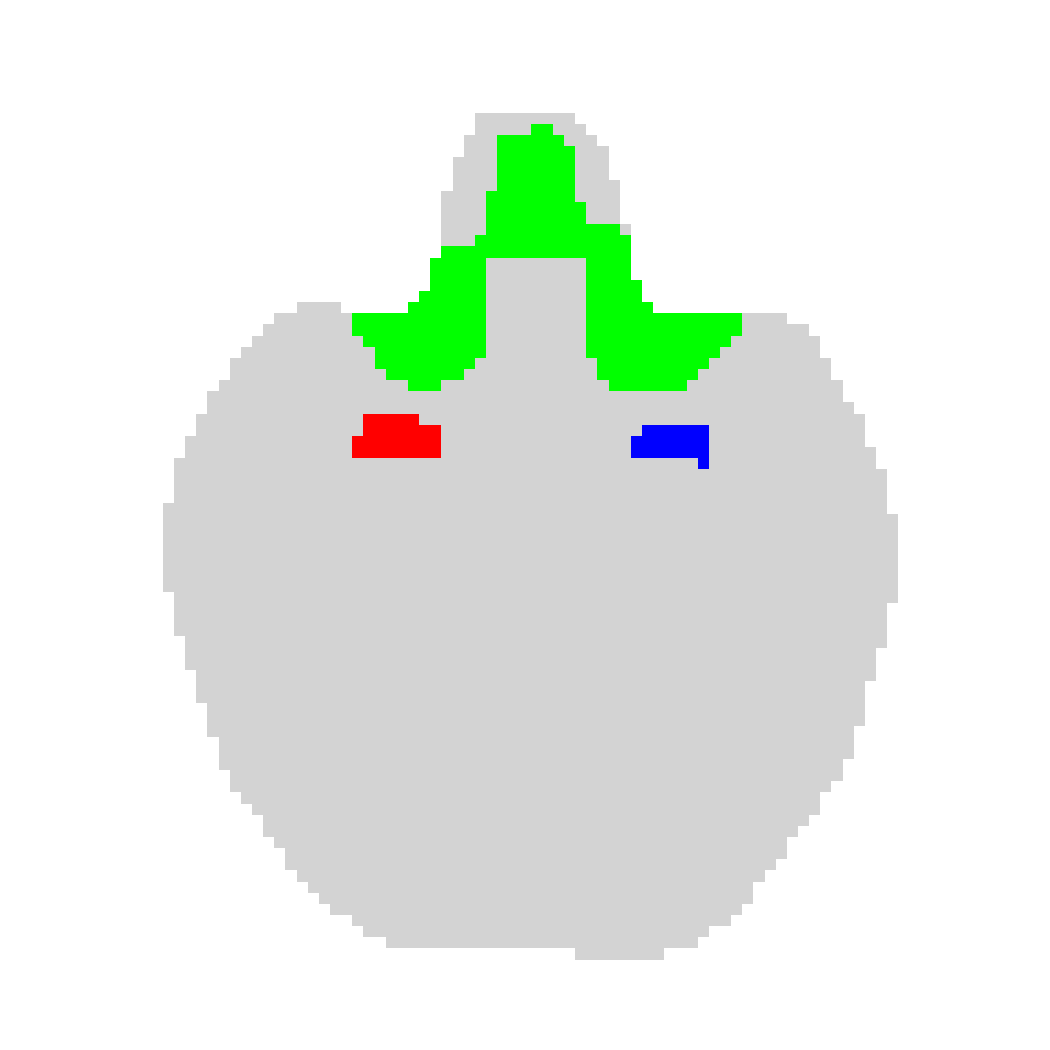}
        \\\vspace{-0.3in}
        \includegraphics[scale=0.30]{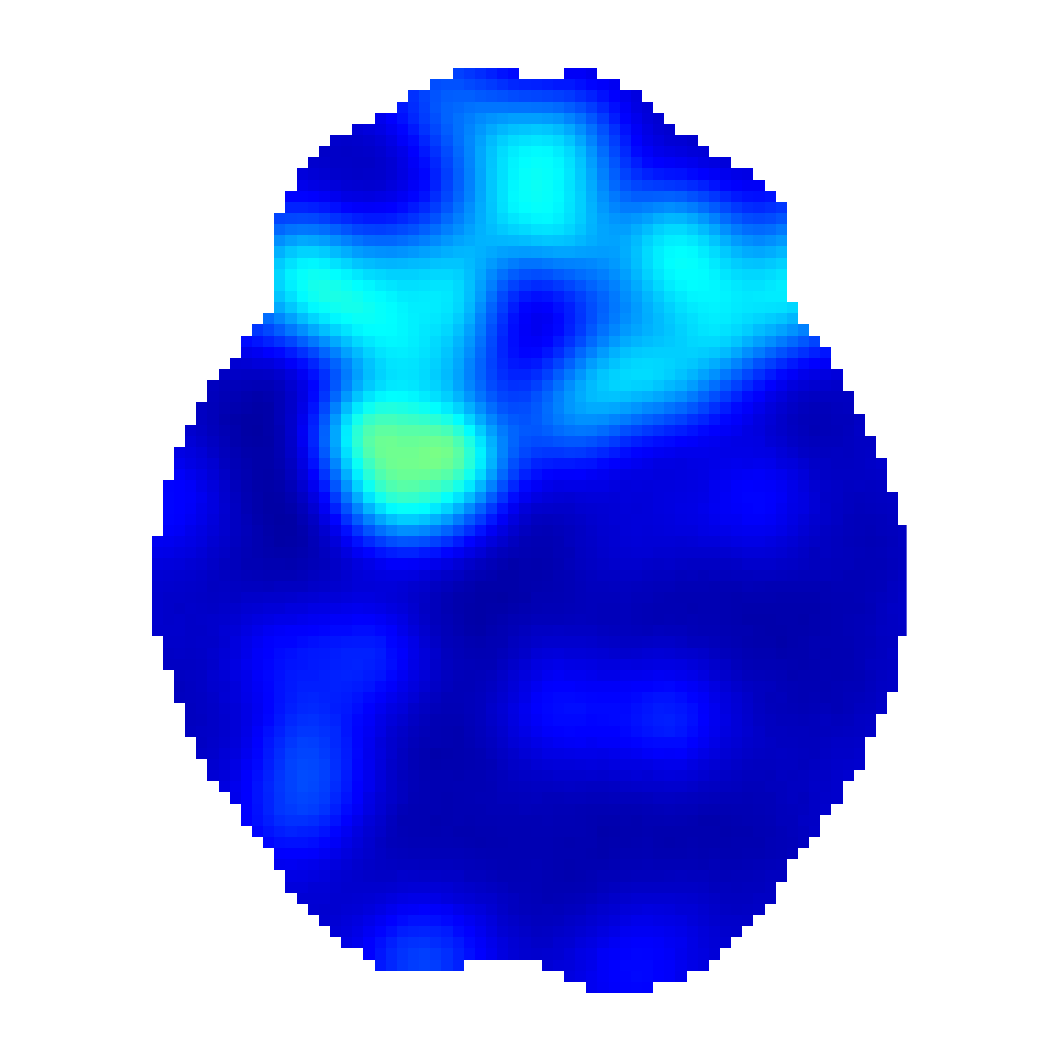}
        \hspace{-0.65in}
        \includegraphics[scale=0.30]{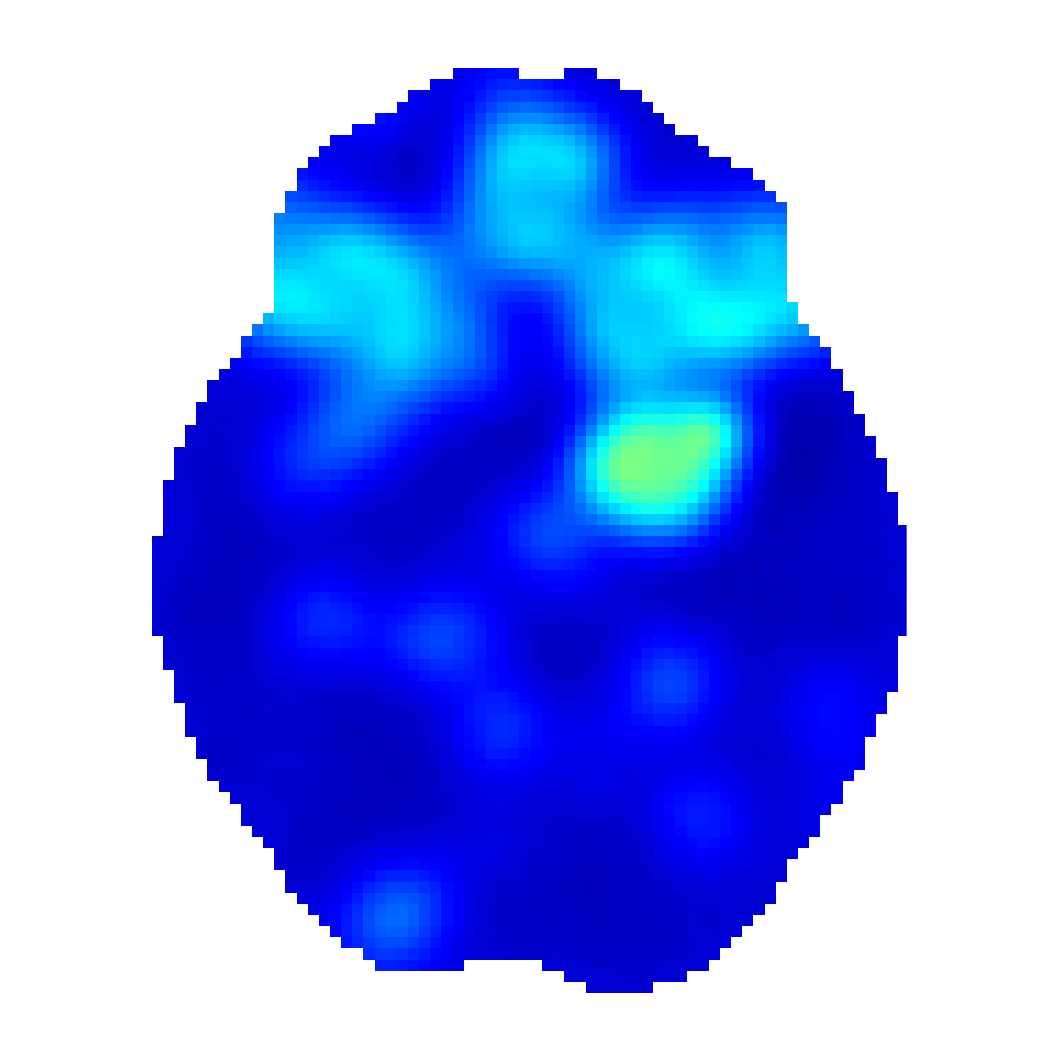}
         \hspace{-0.65in}
        \includegraphics[scale=0.30]{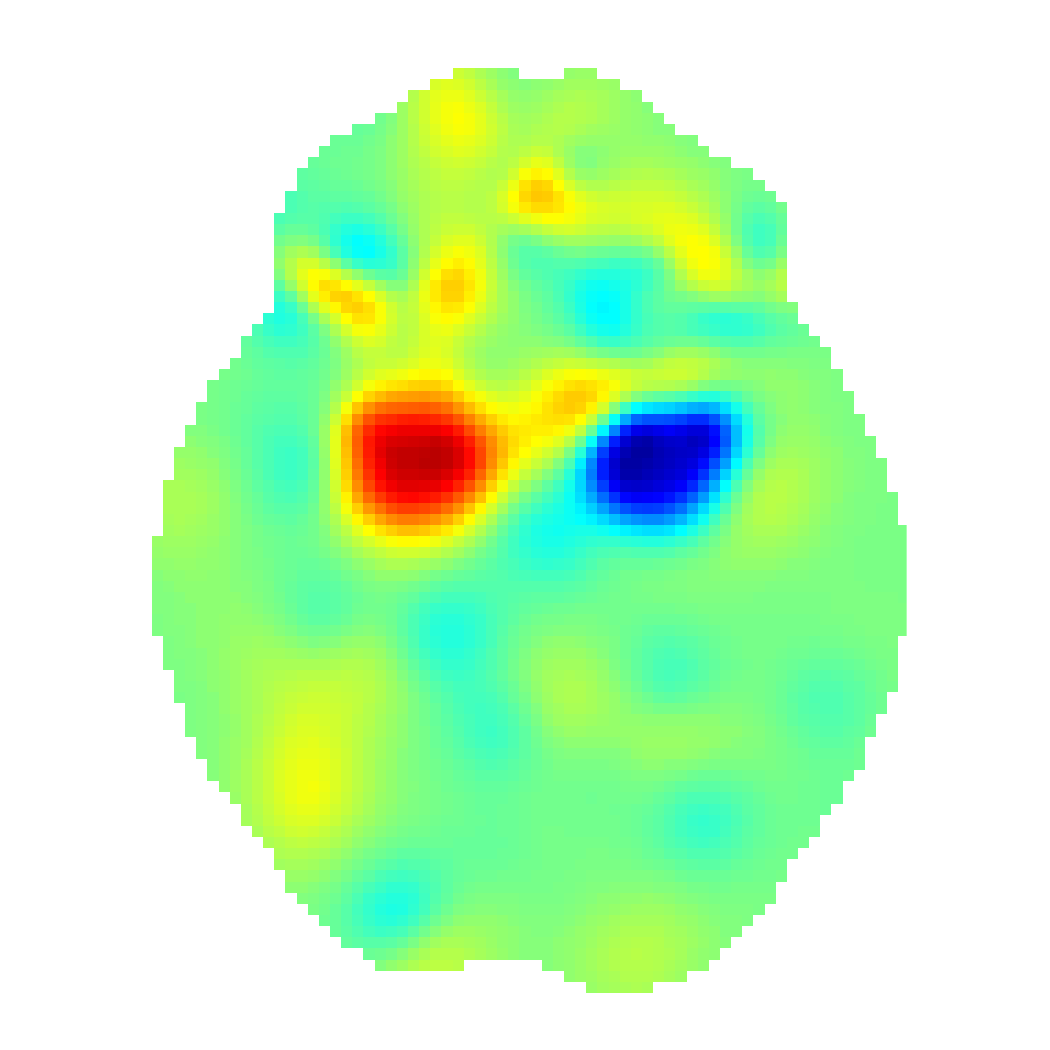}
         \hspace{-0.65in}
        \includegraphics[scale=0.30]{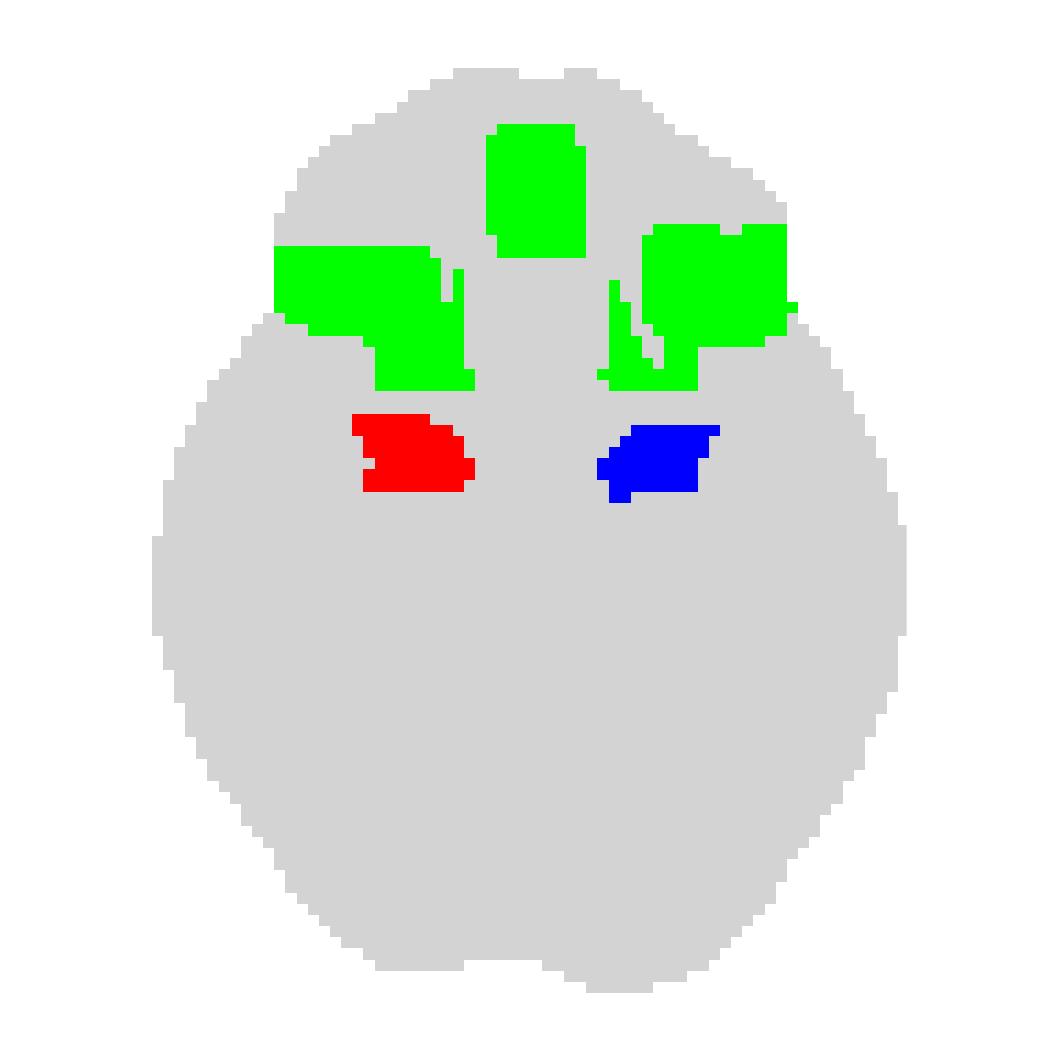}\\ 
         \vspace{-0.1in}
         \hspace{-3.15in}
         \includegraphics[scale=0.4]{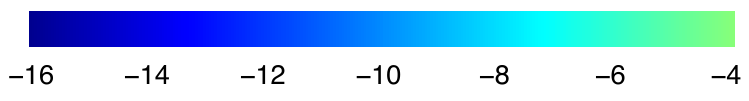}
        \vspace{-0.01in}
      \caption{Results for simulation setup 2 of Section \ref{subsec:simII}. The top row corresponds to axial slice $z=-24$ whereas the bottom row corresponds to axial slice $z=-16$. Columns 1 and 2 are the estimated log-intensities for type 1 and type 2 respectively. The third column is the standardised mean posterior difference between the two latent Gaussian processes in the corresponding slice; bright colours indicate areas mostly activated by type 1 process. The fourth column shows the regions of the brain systematically activated by the two processes; red for type 1, blue for type 2 and green for both.}      \label{fig:sim2comb}
 \end{figure}
\end{landscape}

In this setup we create datasets with a pattern of points that follows brain structures of interest. 
Again there are two types of studies, say type 1 and type 2. 
For each study $i$, $i=1,\dots,200$, we generate the total number of points from a Negative Binomial distribution with mean $\mu=6+2z_{i3}-\mathbf{1}_{\left\{z_{i4}=0\right\}}+\mathbf{1}_{\left\{z_{i4}=1\right\}}$ and variance $\mu^2/20$. 
For the covariates, $z_{i3}\sim\mathrm{Uni[-1,1]}$ and $z_{i4}\sim\mathrm{Bernoulli}(0.5)$. 
Once we know the exact number of foci per study, we assign the study uniformly at random to one of the 2 types and the distribute its foci as follows.  
For type 1, foci appear systematically in the following regions: each focus can be observed in the right amygdala ($B_R$) with probability 55\%, the orbifrontal cortex ($B_C$) with probability 30\% or anywhere else in the brain with probability 15\%. 
The configuration for type 2 differs in that most of the points will go to the left amygdala ($B_L$) instead of the right amygdala. 
If a focus is assigned to one of the three broad regions, the exact location has a uniform distribution over the region. 
In the fourth column of Figure \ref{fig:sim2comb} the regions in red and blue correspond to the left and right amygdala respectively while the orbifrontal cortex is coloured in green.

HMC is run for 10,000 iterations, discarding the 4,000 first as a burn-in and saving every 6 to obtain a total of 1,000 samples from the posterior.  
The run took approximately 15 hours on a Tesla K20c GPU card.

Results are shown in Figure \ref{fig:sim2comb} where in the first two columns we see median posterior log-intensities for the two types, in different axial slices. 
In both cases, we find that the regions with the highest intensities are the amygdalae and that the orbifrontal cortex is a region of high posterior intensity as well. 
The median expected number of points is 5.81 for type 1 (95\% CI [5.36,6,32]) and 6.45 for type 2 (95\% CI [5.97,6.97]). 
The observed values are 6.27 and 6.73 respectively. 

Conditional on there being exactly one focus, we can estimate the probability that this focus appears in any subset $B\subseteq\mathcal{B}$ as $\int_{B}{\lambda\left(\xi\right)d\xi}/\int_{\mathcal{B}}{\lambda\left(\xi\right)d\xi}$. 
Using the posterior draws obtained from the HMC algorithm, we can obtain the posterior distribution of any such quantity. 
For our simulated type 1 data we find that the median posterior probability of observing a focus in the right amygdala ($B_R$) is 0.43 (95\% CI [0.40,0.48]).
For type 2, the probability of observing a focus in the left amygdala ($B_L$) is 0.42 (95\% CI [0.39,0.46]).
For the orbifrontal cortex  ($B_C$) the median posterior probabilities are 0.25 for type 1 and 0.23 for type 2, with 95\% credible intervals $[0.22,0.28]$ and $[0.20,0.26]$ respectively. 
We therefore see that the model underestimates the probabilities for $B_R$, $B_L$ and $B_C$. 
This bias can be attributed to the smoothness that is imposed by our parameter $\delta$ thus leading to increases intensities just outside these regions as well as regions where noise foci appear.   

An interesting question one may ask is which are the regions of the brain that are activated by one type or the other, but not both. 
To answer this, one can construct the mean standardised posterior difference map computed as the ratio of the posterior mean of the difference $\left(\boldsymbol\beta_1\right)_v-\left(\boldsymbol\beta_2\right)_v$, to the posterior standard deviation of that difference: 
$\frac{\left(\boldsymbol\beta_1\right)_v-\left(\boldsymbol\beta_2\right)_v}{\mathrm{sd}\left(\left(\boldsymbol\beta_1\right)_v-\left(\boldsymbol\beta_2\right)_v\right)}$. 
Extreme negative or positive values are evidence of differences between the two types. 
We show the difference map in the third column of Figure \ref{fig:sim2comb}. 
As we see, the model distinguishes the the two types in the amygdala but the differences are small in the rest of the brain.

\newpage
\section{Analysis of the WM dataset: supplementary plots}\label{sec:datasupp}
\begin{landscape}
\begin{figure}[h]
		\centering
         \includegraphics[scale=1.30]{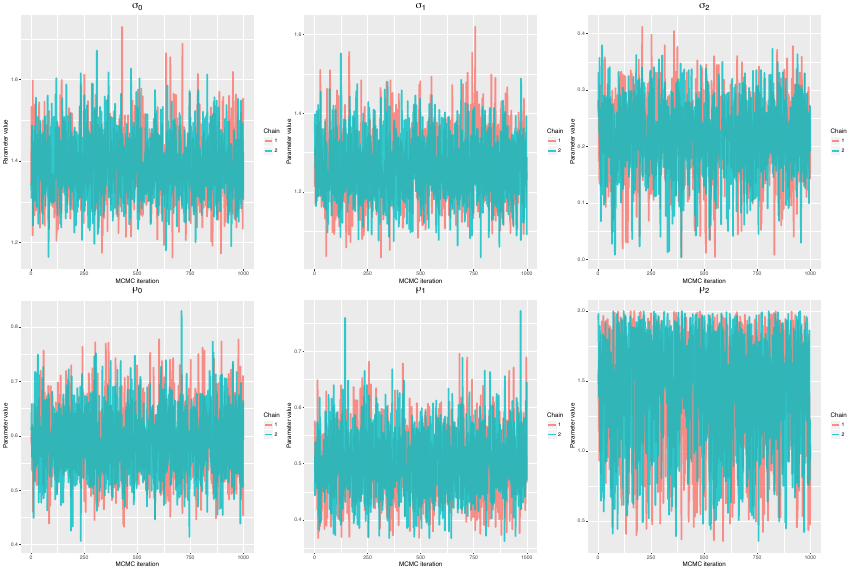}
       \caption{Posterior traceplots for some of the scalar parameters of the LGCP model of Equation \ref{eq:reallgcp} that we fit to the working memory dataset presented in Section \ref{sec:dataintro}. Top row: marginal standard deviations $\sigma_k$ of the 3 latent GPs. Bottom row: correlation decay parameters $\rho_k$ of the 3 latent GPs. The chains have been obtained after applying a thinning factor of 15 to the original chains of length 15,000.}
      \label{fig:trace1}
 \end{figure}
\end{landscape}

\begin{landscape}
\begin{figure}[h]
		\centering
         \includegraphics[scale=1.30]{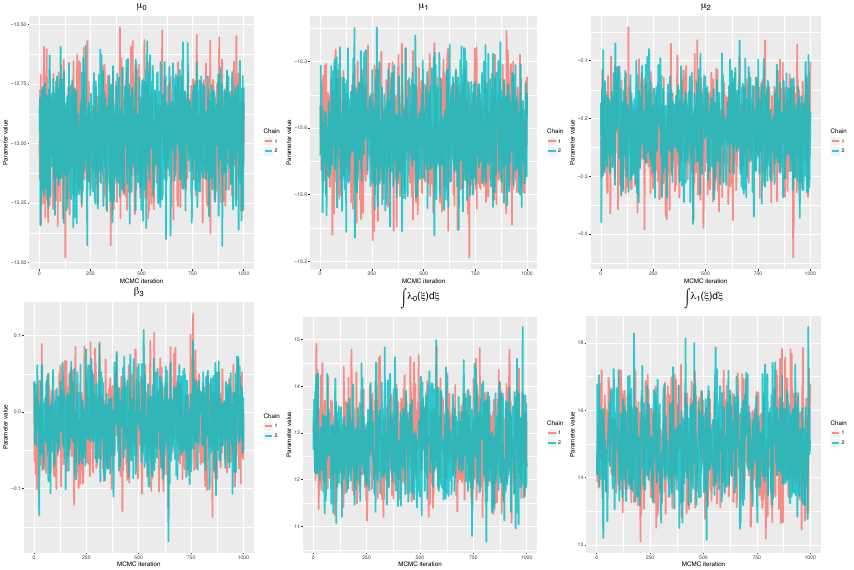}
       \caption{Posterior traceplots for some of the scalar parameters of the LGCP model of Equation \ref{eq:reallgcp} that we fit to the working memory dataset presented in Section \ref{sec:dataintro}. Top row: overall means $\mu_k$ of the 3 latent GPs. Bottom row, subplot 1: regression coefficient $\beta_3$ for the sample size covariate. Bottom row, subplots 2-3: integrated intensities for verbal and non-verbal studies, respectively. The chains have been obtained after applying a thinning factor of 15 to the original chains of length 15,000.}
      \label{fig:trace2}
 \end{figure}
\end{landscape}

\begin{landscape}
\begin{figure}[h]
		\centering
         \includegraphics[scale=1.30]{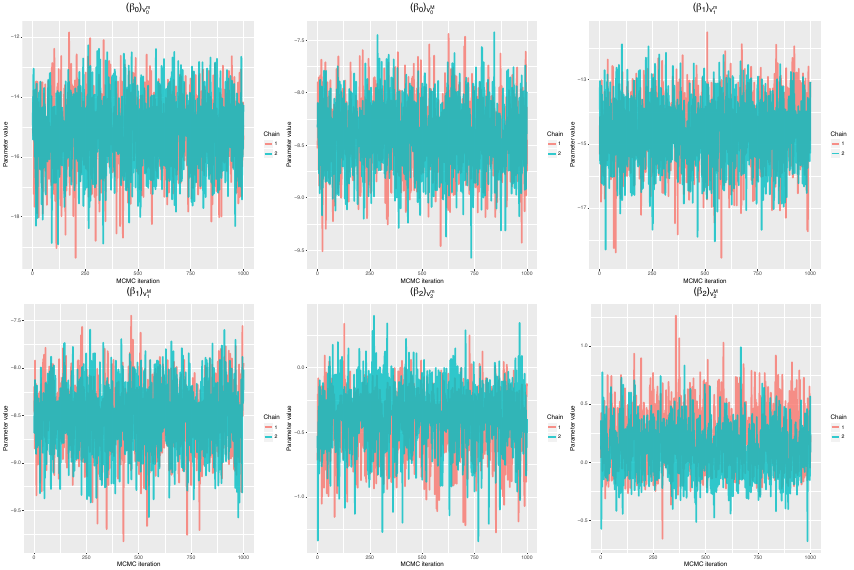}
       \caption{Posterior traceplots for some of the scalar parameters of the LGCP model of Equation \ref{eq:reallgcp} that we fit to the working memory dataset presented in Section \ref{sec:dataintro}. Subplots 1-6 correspond to $\left(\boldsymbol{\beta}_0\right)_{v_0^m}$, $\left(\boldsymbol{\beta}_0\right)_{v_0^M}$, $\left(\boldsymbol{\beta}_1\right)_{v_1^m}$, $\left(\boldsymbol{\beta}_1\right)_{v_1^M}$, $\left(\boldsymbol{\beta}_2\right)_{v_2^m}$ and $\left(\boldsymbol{\beta}_2\right)_{v_2^M}$, where $v_k^m$ and  $v_k^M$ represent the voxels with the minimum and maximum mean posterior values of $\boldsymbol{\beta}_k$, respectively. The chains have been obtained after applying a thinning factor of 15 to the original chains of length 15,000.}
      \label{fig:trace3}
 \end{figure}
\end{landscape}

\begin{figure}[h]
		\centering
         \includegraphics[scale=1.30]{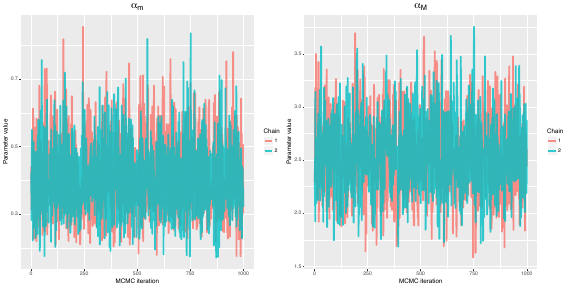}
       \caption{Posterior traceplots for some of the scalar parameters of the LGCP model of Equation \ref{eq:reallgcp} that we fit to the working memory dataset presented in Section \ref{sec:dataintro}. Subplots 1-2 represent $\alpha_m$ and $\alpha_M$, where $m$ and $M$ index the studies with the lowest and largest mean posterior random effects, respectively. The chains have been obtained after applying a thinning factor of 15 to the original chains of length 15,000.}
      \label{fig:trace4}
 \end{figure}

\begin{landscape}
\begin{figure}[htp]
		\centering
        \vspace{-0.90in}
        \includegraphics[scale=1.05]{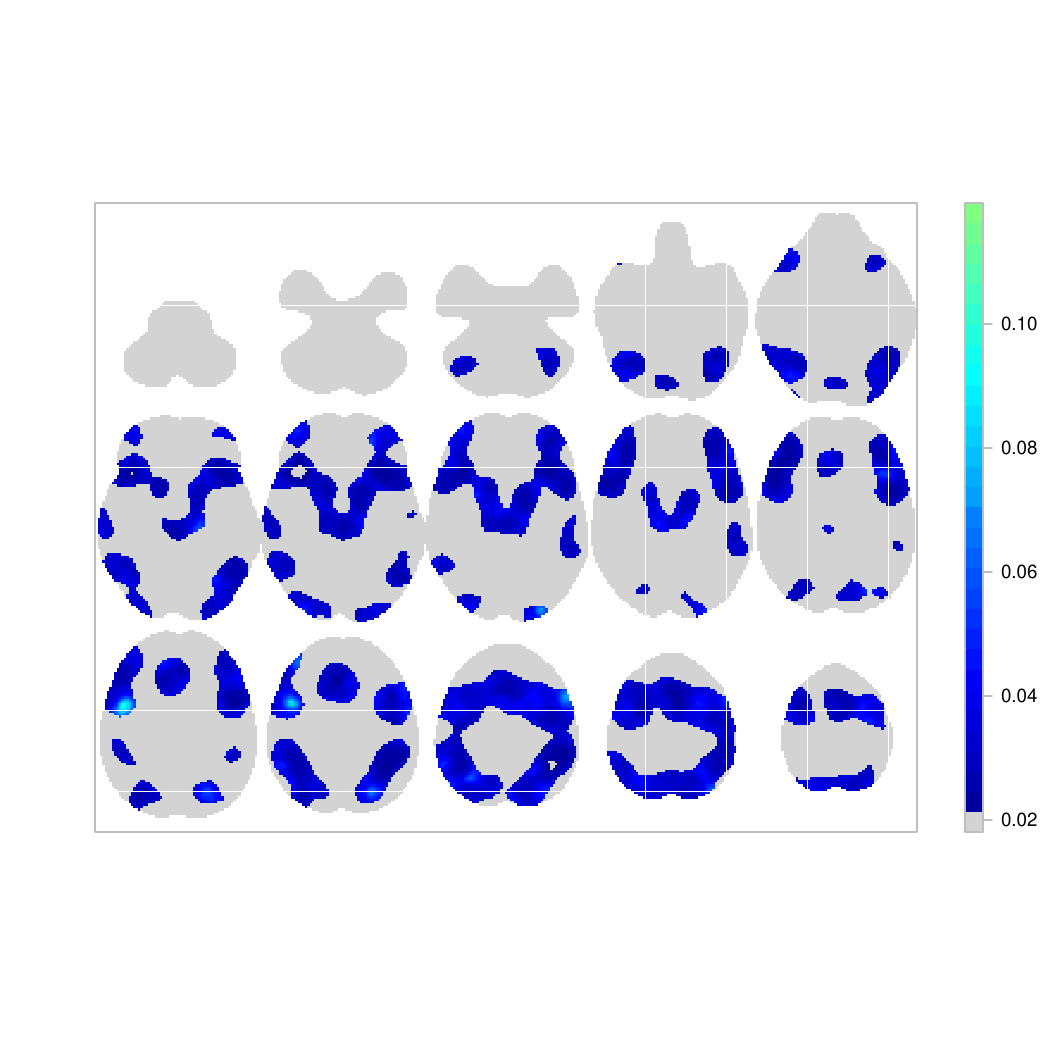}
        \vspace{-1.4in}
       \caption{Posterior variance of $\exp{\left\{\boldsymbol{\beta}_2\right\}}$, the multiplicative age effect on the intensity of both verbal and non-verbal studies. Top row shows (from left to right) axial slices $z=-50,-42,-34,-26$ and $-18$, respectively. Middle row shows axial slices $z=-10,-2,+6,+14$ and $+22$, respectively. Bottom row shows axial slices $z=+30,+38,+46,+54$ and $+62$, respectively. Voxels for which the mean posterior $\boldsymbol\lambda$ is low (below the 75\% quantile over the brain) have been set to zero.}
      \label{fig:agevar}
 \end{figure}
\end{landscape}

This section contains MCMC convergence diagnostics and supplemental material for the real data analysis of Section \ref{sec:data}. 
Posterior traceplots are shown in Figures \ref{fig:trace1}, \ref{fig:trace2}, \ref{fig:trace3} and \ref{fig:trace4}.
The posterior traceplots for the marginal standard deviation parameters $\sigma_k$ and the correlation decay parameters $\rho_k$ are shown in Figure \ref{fig:trace1} (top and bottom row, respectively). 
Figure \ref{fig:trace2} shows posterior traceplots for the overall mean parameters $\mu_k$ (top row), the regression coefficient of the sample size covariate (bottom row, first subplot) and the integrated intensities of verbal and non-verbal studies (bottom row, second and third subplot, respectively). 
Let $v_k^m$ and $v_k^M$ be the voxels where the $k$-th latent GP $\boldsymbol\beta_k$ has the minimum and maximum mean posterior values, respectively. 
Posterior traceplots of $\left(\boldsymbol{\beta}_0\right)_{v_0^m}$, $\left(\boldsymbol{\beta}_0\right)_{v_0^M}$, $\left(\boldsymbol{\beta}_1\right)_{v_1^m}$, $\left(\boldsymbol{\beta}_1\right)_{v_1^M}$, $\left(\boldsymbol{\beta}_2\right)_{v_2^m}$ and $\left(\boldsymbol{\beta}_2\right)_{v_2^M}$ are shown in Figure \ref{fig:trace3}. 
Finally, in Figure \ref{fig:trace4} we present $\alpha_m$ and $\alpha_M$, where $m$ and $M$ index the studies with the lowest and largest mean posterior random effects, respectively. 
All chains are obtained after applying a thinning factor of 15 to the original MCMC chains of length 15,000. 
Finally, Figure \ref{fig:agevar} shows the voxelwise posterior variance of $\exp{\left\{\boldsymbol{\beta}_2\right\}}$, the multiplicative age effect on the intensity of both verbal and non-verbal studies.

\newpage
\section{Full brain analysis}\label{app:fullbrain}
In this section we present full brain results for the working memory CBMA conducted in Section \ref{sec:data}. 
Table \ref{tab:fullbrain} presents probabilities of activation and expected number of foci for several ROIs, along with their 95\% credible intervals. 
In Table \ref{tab:fullbrain2} we compare the probability of an activation between working memory studies using verbal and non-verbal stimuli, using the same ROIs. 
The ROIs have been obtained from the Harvard-Oxford atlas \citep{Desikan2006}. 
All quantities are based on 1,000 MCMC samples which are obtained after applying a thinning factor of 15 to the original MCMC chains of length 15,000.

\begin{landscape}
\begin{longtable}{lr|rrr|rrr}
\caption{Working memory CBMA results. Table presents posterior summaries for the \% probability of at least one activation in a given ROI (rounded to 2 decimal points), as well as the expected number of foci (rounded to 3 decimal points). All quantities have been calculated based on 1,000 MCMC samples.}
\label{tab:fullbrain}\\ 
 & & \multicolumn{3}{c}{$\mathbf{\mathbb{P}\left(N_X(B)\geq 1\right)}$} & \multicolumn{3}{c}{$\mathbf{\int_B\lambda(\xi)d\xi}$}\\
\textbf{ROI} & \textbf{Vol.} & \textbf{Mean} & $\mathbf{p_{0.025}}$ & $\mathbf{p_{0.975}}$ & \textbf{Mean} & $\mathbf{p_{0.025}}$ & $\mathbf{p_{0.975}}$\\ 
\hline\hline
Frontal pole & 25900 & 60.24 & 51.05 & 66.42 & 0.926 & 0.714 & 1.091 \\ 
  Insular cortex & 3613 & 33.39 & 26.68 & 39.36 & 0.407 & 0.310 & 0.500 \\ 
  Superior frontal gyrus & 8861 & 38.27 & 27.74 & 43.98 & 0.484 & 0.325 & 0.579 \\ 
  Middle frontal gyrus & 8421 & 60.55 & 48.23 & 66.48 & 0.933 & 0.658 & 1.093 \\ 
  Inferior frontal gyrus, pars triangularis & 2317 & 22.54 & 15.44 & 27.94 & 0.256 & 0.168 & 0.328 \\ 
  Inferior frontal gyrus, pars opercularis & 2335 & 39.88 & 31.06 & 45.96 & 0.510 & 0.372 & 0.615 \\ 
  Precentral gyrus & 13967 & 68.47 & 59.96 & 73.72 & 1.158 & 0.915 & 1.336 \\ 
  Temporal pole & 8044 & 9.60 & 5.41 & 13.48 & 0.101 & 0.056 & 0.145 \\ 
  Superior temporal gyrus, anterior division & 916 & 1.77 & 0.80 & 2.87 & 0.018 & 0.008 & 0.029 \\ 
  Superior temporal gyrus, posterior division & 2897 & 8.89 & 5.46 & 11.91 & 0.093 & 0.056 & 0.127 \\ 
  Middle temporal gyrus, anterior division & 1425 & 1.64 & 0.55 & 2.88 & 0.017 & 0.005 & 0.029 \\ 
  Middle temporal gyrus, posterior division & 4206 & 9.37 & 5.18 & 12.76 & 0.099 & 0.053 & 0.137 \\ 
  Middle temporal gyrus, temporooccipital part & 3202 & 12.84 & 6.85 & 17.63 & 0.138 & 0.071 & 0.194 \\ 
  Inferior temporal gyrus, anterior division & 1287 & 1.38 & 0.38 & 2.56 & 0.014 & 0.004 & 0.026 \\ 
  Inferior temporal gyrus, posterior division & 4138 & 5.43 & 2.80 & 8.12 & 0.056 & 0.028 & 0.085 \\ 
  Inferior temporal gyrus, temporooccipital part & 2605 & 11.49 & 6.16 & 16.05 & 0.122 & 0.064 & 0.175 \\ 
  Postcentral gyrus & 10638 & 24.29 & 17.60 & 29.24 & 0.279 & 0.194 & 0.346 \\ 
  Superior parietal lobule & 4489 & 36.16 & 26.16 & 42.31 & 0.450 & 0.303 & 0.550 \\ 
  Supramarginal gyrus, anterior division & 2910 & 13.19 & 7.98 & 17.24 & 0.142 & 0.083 & 0.189 \\ 
  Supramarginal gyrus, posterior division & 4071 & 28.38 & 19.69 & 33.72 & 0.335 & 0.219 & 0.411 \\ 
  Angular gyrus & 3703 & 21.69 & 14.39 & 26.34 & 0.245 & 0.155 & 0.306 \\ 
  Lateral occipital cortex, superior division & 14484 & 61.52 & 53.01 & 67.73 & 0.959 & 0.755 & 1.131 \\ 
  Lateral occipital cortex, inferior division & 7490 & 27.53 & 18.66 & 33.57 & 0.323 & 0.207 & 0.409 \\ 
  Intracalcarine cortex & 2211 & 8.36 & 4.35 & 11.76 & 0.087 & 0.045 & 0.125 \\ 
  Frontal medial cortex & 1539 & 1.94 & 0.64 & 3.62 & 0.020 & 0.006 & 0.037 \\ 
  Juxtapositional lobule cortex & 2282 & 26.46 & 16.14 & 32.74 & 0.308 & 0.176 & 0.397 \\ 
  Subcallosal cortex & 2176 & 4.47 & 1.73 & 7.16 & 0.046 & 0.017 & 0.074 \\ 
  Paracingulate gyrus & 4095 & 46.22 & 35.94 & 52.89 & 0.622 & 0.445 & 0.753 \\ 
  Cingulate gyrus, anterior division & 4144 & 22.61 & 14.76 & 28.19 & 0.257 & 0.160 & 0.331 \\ 
  Cingulate gyrus, posterior division & 4668 & 11.07 & 5.70 & 15.52 & 0.118 & 0.059 & 0.169 \\ 
  Precuneous cortex & 7844 & 26.46 & 18.08 & 32.62 & 0.308 & 0.199 & 0.395 \\ 
  Cuneal cortex & 1743 & 7.75 & 4.09 & 10.95 & 0.081 & 0.042 & 0.116 \\ 
  Frontal orbital cortex & 5188 & 36.94 & 27.27 & 43.06 & 0.462 & 0.318 & 0.563 \\ 
  Parahippocampal gyrus, anterior division & 3313 & 5.69 & 2.94 & 8.43 & 0.059 & 0.030 & 0.088 \\ 
  Parahippocampal gyrus, posterior division & 2014 & 6.01 & 3.23 & 8.71 & 0.062 & 0.033 & 0.091 \\ 
  Lingual gyrus & 5388 & 15.62 & 9.12 & 20.15 & 0.170 & 0.096 & 0.225 \\ 
  Temporal fusiform cortex, anterior division & 1243 & 1.21 & 0.39 & 2.22 & 0.012 & 0.004 & 0.022 \\ 
  Temporal fusiform cortex, posterior division & 2951 & 5.72 & 3.10 & 8.42 & 0.059 & 0.032 & 0.088 \\ 
  Temporal occipital fusiform cortex & 2458 & 13.65 & 8.21 & 17.69 & 0.147 & 0.086 & 0.195 \\ 
  Occipital fusiform gyrus & 3587 & 23.35 & 15.64 & 28.33 & 0.266 & 0.170 & 0.333 \\ 
  Frontal operculum cortex & 1062 & 16.48 & 9.97 & 20.60 & 0.180 & 0.105 & 0.231 \\ 
  Central opercular cortex & 2578 & 11.08 & 6.70 & 14.67 & 0.118 & 0.069 & 0.159 \\ 
  Parietal operculum cortex & 1684 & 6.23 & 3.23 & 8.92 & 0.064 & 0.033 & 0.093 \\ 
  Planum polare & 1210 & 2.64 & 1.29 & 3.94 & 0.027 & 0.013 & 0.040 \\ 
  Heschl's gyrus & 786 & 2.51 & 1.08 & 3.89 & 0.025 & 0.011 & 0.040 \\ 
  Planum temporale & 1442 & 6.05 & 3.26 & 8.45 & 0.063 & 0.033 & 0.088 \\ 
  Supracalcarine cortex & 424 & 1.52 & 0.66 & 2.41 & 0.015 & 0.007 & 0.024 \\ 
  Occipital pole & 9658 & 20.42 & 12.04 & 25.75 & 0.229 & 0.128 & 0.298 \\ 
  Left cerebral white matter & 28034 & 81.33 & 75.05 & 85.13 & 1.684 & 1.388 & 1.906 \\ 
  Left cerebral cortex & 82249 & 99.04 & 98.05 & 99.44 & 4.679 & 3.938 & 5.193 \\ 
  Left lateral ventricle & 1289 & 4.10 & 1.96 & 6.02 & 0.042 & 0.020 & 0.062 \\ 
  Left thalamus & 1591 & 16.17 & 9.31 & 21.64 & 0.177 & 0.098 & 0.244 \\ 
  Left caudate & 572 & 3.60 & 1.75 & 5.44 & 0.037 & 0.018 & 0.056 \\ 
  Left putamen & 923 & 7.83 & 3.91 & 10.96 & 0.082 & 0.040 & 0.116 \\ 
  Left pallidum & 312 & 5.15 & 2.58 & 7.59 & 0.053 & 0.026 & 0.079 \\ 
  Brain stem & 8078 & 13.45 & 7.80 & 18.29 & 0.145 & 0.081 & 0.202 \\ 
  Left hippocampus & 921 & 3.38 & 1.21 & 5.37 & 0.034 & 0.012 & 0.055 \\ 
  Left amygdala & 390 & 1.41 & 0.48 & 2.43 & 0.014 & 0.005 & 0.025 \\ 
  Left accumbens & 111 & 0.88 & 0.26 & 1.57 & 0.009 & 0.003 & 0.016 \\ 
  Right cerebral white matter & 31216 & 79.16 & 72.13 & 83.13 & 1.573 & 1.278 & 1.780 \\ 
  Right cerebral cortex & 86480 & 98.78 & 97.36 & 99.24 & 4.431 & 3.636 & 4.874 \\ 
  Right lateral ventricle & 1019 & 3.36 & 1.77 & 4.97 & 0.034 & 0.018 & 0.051 \\ 
  Right thalamus & 1398 & 12.88 & 7.12 & 17.43 & 0.138 & 0.074 & 0.192 \\ 
  Right caudate & 515 & 2.68 & 1.02 & 4.30 & 0.027 & 0.010 & 0.044 \\ 
  Right putamen & 800 & 6.82 & 3.39 & 9.80 & 0.071 & 0.034 & 0.103 \\ 
  Right pallidum & 266 & 3.24 & 1.08 & 5.19 & 0.033 & 0.011 & 0.053 \\ 
  Right hippocampus & 772 & 1.78 & 0.67 & 2.99 & 0.018 & 0.007 & 0.030 \\ 
  Right amygdala & 399 & 0.86 & 0.19 & 1.73 & 0.009 & 0.002 & 0.017 \\ 
  Right accumbens & 86 & 0.64 & 0.15 & 1.26 & 0.006 & 0.001 & 0.013 \\ 
  \hline
\end{longtable} 
\end{landscape}

\begin{landscape}
\begin{longtable}{lr|rrr|rrr}
\caption{Working memory CBMA results. Table presents posterior summaries for the \% probability of a reported activation in studies using verbal and non-verbal stimuli. All quantities have been calculated based on 1,000 MCMC samples.}
\label{tab:fullbrain2}\\ 
& & \multicolumn{3}{c}{\textbf{Verbal}} & \multicolumn{3}{c}{\textbf{Non-verbal}}\\
\textbf{ROI} & \textbf{Vol.} & \textbf{Mean} & $\mathbf{p_{0.025}}$ & $\mathbf{p_{0.975}}$ & \textbf{Mean} & $\mathbf{p_{0.025}}$ & $\mathbf{p_{0.975}}$\\ 
\hline\hline
Frontal pole & 25900 & 59.99 & 46.81 & 67.86 & 60.24 & 41.75 & 69.92 \\ 
  Insular cortex & 3613 & 32.79 & 21.14 & 40.70 & 33.86 & 20.43 & 43.28 \\ 
  Superior frontal gyrus & 8861 & 29.45 & 18.36 & 36.70 & 45.87 & 29.60 & 54.29 \\ 
  Middle frontal gyrus & 8421 & 53.05 & 42.42 & 60.85 & 66.66 & 50.91 & 74.55 \\ 
  Inferior frontal gyrus, pars triangularis & 2317 & 19.43 & 10.84 & 24.95 & 25.44 & 14.64 & 33.90 \\ 
  Inferior frontal gyrus, pars opercularis & 2335 & 43.66 & 31.68 & 51.15 & 35.69 & 21.49 & 44.80 \\ 
  Precentral gyrus & 13967 & 64.10 & 51.56 & 71.00 & 72.09 & 57.82 & 79.52 \\ 
  Temporal pole & 8044 & 7.54 & 2.91 & 11.50 & 11.58 & 4.28 & 18.15 \\ 
  Superior temporal gyrus, anterior division & 916 & 2.31 & 0.71 & 4.05 & 1.23 & 0.27 & 2.52 \\ 
  Superior temporal gyrus, posterior division & 2897 & 10.76 & 5.31 & 15.14 & 6.95 & 2.25 & 11.32 \\ 
  Middle temporal gyrus, anterior division & 1425 & 2.19 & 0.48 & 4.36 & 1.09 & 0.15 & 2.56 \\ 
  Middle temporal gyrus, posterior division & 4206 & 9.58 & 4.59 & 14.22 & 9.12 & 3.83 & 14.44 \\ 
  Middle temporal gyrus, temporooccipital part & 3202 & 11.88 & 5.65 & 17.75 & 13.73 & 5.73 & 21.06 \\ 
  Inferior temporal gyrus, anterior division & 1287 & 1.76 & 0.31 & 3.69 & 0.99 & 0.10 & 2.53 \\ 
  Inferior temporal gyrus, posterior division & 4138 & 4.85 & 1.59 & 8.09 & 5.98 & 1.47 & 10.31 \\ 
  Inferior temporal gyrus, temporooccipital part & 2605 & 7.45 & 3.00 & 11.83 & 15.31 & 6.95 & 22.66 \\ 
  Postcentral gyrus & 10638 & 20.37 & 11.48 & 26.10 & 27.95 & 16.64 & 35.68 \\ 
  Superior parietal lobule & 4489 & 38.81 & 26.05 & 46.48 & 33.24 & 19.67 & 42.96 \\ 
  Supramarginal gyrus, anterior division & 2910 & 12.78 & 5.90 & 17.83 & 13.56 & 6.23 & 19.83 \\ 
  Supramarginal gyrus, posterior division & 4071 & 29.93 & 19.40 & 36.61 & 26.70 & 14.58 & 35.17 \\ 
  Angular gyrus & 3703 & 24.30 & 15.51 & 31.09 & 18.91 & 9.32 & 25.61 \\ 
  Lateral occipital cortex, superior division & 14484 & 55.45 & 42.51 & 63.87 & 66.55 & 52.60 & 74.58 \\ 
  Lateral occipital cortex, inferior division & 7490 & 23.69 & 14.54 & 30.51 & 31.06 & 18.71 & 40.58 \\ 
  Intracalcarine cortex & 2211 & 6.51 & 2.61 & 10.12 & 10.15 & 4.37 & 15.19 \\ 
  Frontal medial cortex & 1539 & 1.21 & 0.24 & 2.62 & 2.66 & 0.60 & 5.58 \\ 
  Juxtapositional lobule cortex & 2282 & 21.21 & 12.33 & 28.23 & 31.23 & 18.26 & 41.74 \\ 
  Subcallosal cortex & 2176 & 4.20 & 1.08 & 7.66 & 4.72 & 1.03 & 9.24 \\ 
  Paracingulate gyrus & 4095 & 42.91 & 29.21 & 51.04 & 49.14 & 33.89 & 59.57 \\ 
  Cingulate gyrus, anterior division & 4144 & 17.69 & 10.95 & 24.25 & 27.15 & 14.86 & 36.16 \\ 
  Cingulate gyrus, posterior division & 4668 & 8.06 & 3.60 & 12.42 & 13.93 & 5.98 & 21.66 \\ 
  Precuneous cortex & 7844 & 22.88 & 12.61 & 30.03 & 29.76 & 18.38 & 39.96 \\ 
  Cuneal cortex & 1743 & 6.71 & 1.97 & 10.43 & 8.76 & 3.29 & 14.54 \\ 
  Frontal orbital cortex & 5188 & 37.26 & 25.37 & 45.24 & 36.48 & 24.30 & 45.43 \\ 
  Parahippocampal gyrus, anterior division & 3313 & 5.93 & 2.24 & 9.59 & 5.44 & 1.65 & 9.35 \\ 
  Parahippocampal gyrus, posterior division & 2014 & 5.80 & 2.27 & 9.47 & 6.21 & 2.46 & 10.81 \\ 
  Lingual gyrus & 5388 & 15.61 & 7.56 & 21.30 & 15.56 & 7.52 & 22.72 \\ 
  Temporal fusiform cortex, anterior division & 1243 & 1.52 & 0.33 & 3.13 & 0.90 & 0.18 & 2.11 \\ 
  Temporal fusiform cortex, posterior division & 2951 & 5.26 & 2.06 & 8.46 & 6.17 & 2.13 & 10.49 \\ 
  Temporal occipital fusiform cortex & 2458 & 8.87 & 4.19 & 13.17 & 18.12 & 8.94 & 26.00 \\ 
  Occipital fusiform gyrus & 3587 & 26.68 & 17.03 & 33.74 & 19.78 & 10.47 & 27.65 \\ 
  Frontal operculum cortex & 1062 & 15.27 & 9.08 & 20.31 & 17.64 & 9.10 & 23.87 \\ 
  Central opercular cortex & 2578 & 10.04 & 4.95 & 14.41 & 12.08 & 5.64 & 17.46 \\ 
  Parietal operculum cortex & 1684 & 7.35 & 2.65 & 11.29 & 5.08 & 1.56 & 9.05 \\ 
  Planum polare & 1210 & 2.37 & 0.79 & 4.02 & 2.91 & 0.98 & 5.02 \\ 
  Heschl's gyrus & 786 & 3.07 & 0.99 & 5.47 & 1.94 & 0.34 & 3.79 \\ 
  Planum temporale & 1442 & 7.38 & 3.50 & 10.90 & 4.69 & 1.55 & 8.26 \\ 
  Supracalcarine cortex & 424 & 0.95 & 0.29 & 1.74 & 2.07 & 0.51 & 3.64 \\ 
  Occipital pole & 9658 & 19.84 & 10.66 & 26.91 & 20.91 & 11.10 & 29.20 \\ 
  Left cerebral white matter & 28034 & 76.33 & 66.31 & 81.39 & 85.13 & 75.35 & 89.94 \\ 
  Left cerebral cortex & 82249 & 98.88 & 97.31 & 99.41 & 99.14 & 97.59 & 99.63 \\ 
  Left lateral ventricle & 1289 & 3.12 & 1.30 & 5.16 & 5.05 & 1.65 & 8.64 \\ 
  Left thalamus & 1591 & 14.56 & 7.44 & 20.87 & 17.69 & 7.35 & 26.38 \\ 
  Left caudate & 572 & 3.49 & 1.25 & 5.71 & 3.69 & 1.09 & 6.53 \\ 
  Left putamen & 923 & 6.97 & 2.01 & 11.15 & 8.67 & 2.25 & 13.88 \\ 
  Left pallidum & 312 & 6.26 & 2.38 & 10.29 & 4.01 & 1.24 & 7.29 \\ 
  Brain stem & 8078 & 10.42 & 5.26 & 15.72 & 16.33 & 6.90 & 25.06 \\ 
  Left hippocampus & 921 & 3.46 & 1.08 & 6.28 & 3.29 & 0.74 & 6.39 \\ 
  Left amygdala & 390 & 1.21 & 0.16 & 2.59 & 1.60 & 0.25 & 3.37 \\ 
  Left accumbens & 111 & 1.07 & 0.15 & 2.29 & 0.69 & 0.09 & 1.49 \\ 
  Right cerebral white matter & 31216 & 73.44 & 65.11 & 78.96 & 83.50 & 74.00 & 88.46 \\ 
  Right cerebral cortex & 86480 & 98.22 & 96.50 & 98.97 & 99.12 & 97.60 & 99.61 \\ 
  Right lateral ventricle & 1019 & 2.66 & 0.96 & 4.35 & 4.05 & 1.21 & 6.74 \\ 
  Right thalamus & 1398 & 15.17 & 8.00 & 21.79 & 10.48 & 3.36 & 16.86 \\ 
  Right caudate & 515 & 2.51 & 0.76 & 4.45 & 2.85 & 0.68 & 5.34 \\ 
  Right putamen & 800 & 5.91 & 1.94 & 9.49 & 7.69 & 2.56 & 12.98 \\ 
  Right pallidum & 266 & 2.74 & 0.65 & 4.97 & 3.73 & 0.93 & 6.91 \\ 
  Right hippocampus & 772 & 1.77 & 0.52 & 3.42 & 1.78 & 0.46 & 3.55 \\ 
  Right amygdala & 399 & 0.86 & 0.11 & 2.03 & 0.87 & 0.13 & 2.17 \\ 
  Right accumbens & 86 & 0.78 & 0.13 & 1.84 & 0.50 & 0.08 & 1.30 \\ 
   \hline
\end{longtable}
\end{landscape}

\end{document}